\documentclass[%
    lettersize,%
    journal,%
]{IEEEtran}

\usepackage{amsmath}
\usepackage{amsfonts}
\usepackage{mathtools}
\usepackage{amsthm}
    \newtheorem{theorem}{Theorem}
\usepackage{dsfont}
\usepackage{algorithmic}
\usepackage{array}
\usepackage[caption = false]{subfig}
\usepackage{textcomp}
\usepackage{stfloats}
\usepackage{url}
\usepackage{cite}
\usepackage{verbatim}
\usepackage{multirow}
\usepackage{graphicx}
\usepackage{balance}
\usepackage{booktabs}
\usepackage{pgf}
\usepackage{pgfplots}
\usepackage{import}
\usepackage{siunitx}
\begin{document}
\title{A Safe and Efficient Control Strategy for a Weakly Instrumented Wheeled Loader}
\author{Louis Roullier, Thomas Chevet, \IEEEmembership{Member, IEEE}, Nicolas Hoffmann, Nicolas Langlois%
    \thanks{%
        This work was partially funded by Association Nationale de la Recherche et de la Technologie and by Banque Publique d'Investissement, R\'egion Normandie, and the European Union through the PROMET\'E project.
        For this reason and the purpose of Open Access, the authors have applied a CC BY public copyright licence to any Author Accepted Manuscript (AAM) version arising from this submission.%
    }%
    \thanks{L. Roullier, T. Chevet, and N. Langlois are with Univ Rouen Normandie, ESIGELEC, Normandie Univ, IRSEEM UR4353, F-76000 Rouen, France (email: \texttt{<}first name\texttt{>}.\texttt{<}last name\texttt{>}@irseem-esigelec.fr).}%
    \thanks{L. Roullier and N. Hoffmann are with HERACLES Robotics, 76000, Rouen, France (email: \texttt{<}first name\texttt{>}.\texttt{<}last name\texttt{>}@heracles-robotics.com).}
}

\maketitle

\begin{abstract}
    Automating heavy machines such as wheeled loaders requires tracking reference trajectories precisely while satisfying safety constraints.
    Some approaches address it, such as model predictive control or safety filters.
    However, they rely on accurate system models that can be hard to obtain especially if the system is weakly-instrumented and its states partially measured.
    This article tackles this problem for a weakly instrumented 23.2-ton articulated wheeled loader.
    The goal is to track a reference trajectory while satisfying constraints on every parts of the machine.
    We propose a hierarchical architecture in which a reinforcement learning (RL) agent, trained offline on a learned model of the machine, tracks the speed commands at the low-level.
    Since it absorbs the complex dynamics, the resulting closed-loop behavior can be predicted by an inexpensive analytic model.
    At the high-level, a nominal policy generates reference commands for the low-level RL agent, which are then corrected by the safety filter if necessary.
    The filter we propose brings both bodies of the machine back onto the reference trajectory after a safety intervention.
    In simulation, it safely avoids every obstacle and stays closer to the reference than existing filters.
    On the real machine, the low-level agent tracks the speed commands accurately while keeping the engine commands smooth.
    Under production conditions, the full architecture runs in real time and keeps the loader both precise and safe.
\end{abstract}


\section{Introduction}

Wheeled loaders are versatile two-articulated-bodies machines used to handle and transport materials of different types (e.g. soil, gravel, etc).
Results in \cite{frank2012introwheelloader} show that their real-world performance in terms of productivity and energy efficiency is strongly influenced by operators’ behaviour.
The automation of wheeled loaders has become an active research topic as it offers the potential to achieve more consistent and optimised machine operation despite environmental uncertainty.
This paper addresses the three main challenges \cite{dadhich2016surveywheelloader,nguyen2023surveywheelloader, feng2025newsurveywheelloader} : precise navigation, complex dynamics, and safety.

For several years, a wide range of controllers have been investigated to automate construction machines.
Recent approaches are mainly based on model predictive control (MPC) or reinforcement learning applied to excavators \cite{egli2021rlexcavatorclassic,ishiara2020excavatorMPC}.
On wheeled loaders, RL has been used to control the scooping \cite{azulay2021rlwheelloader}.
MPC has also been applied to articulated vehicles.
In \cite{nayl2012wheelloadermodel}, a switching MPC tracks a path with an articulated vehicle under varying slip angles, in \cite{Penttila2025wheelloadermpcrl} an actor-critic MPC navigates a wheeled loader towards a goal.
Furthermore, \cite{majd20254mppitractor} uses model predictive path integral (MPPI) controller accelerated on GPU to control a tractor-trailer under safety constraints encoded by barrier functions.

These methods require a model of the machine which is difficult to obtain.
Internal events, such as automatic gear shifts, may not be observed.
A learned model can reproduce the dynamics of such machines from data \cite{hoffmann2025lstmexcavator}.
However, using this model online is costly.
MPPI samples many trajectories of the model at each time step \cite{asmar2023mppiadaptiveimportancesampling,keshavarz2025mppice}.
Although learned models can be incorporated into MPC, their integration may introduce non-convexities into the optimization problem.
In \cite{lahr2026l4acados}, the neural network is linearized and evaluated on a GPU to improve computational efficiency.
Nevertheless, the computational cost remains dependent on both the prediction horizon and model complexity.
Beyond computational considerations, another important distinction between MPC and MPPI lies in their treatment of constraints.
MPC can explicitly enforce both linear and nonlinear constraints within the optimization problem \cite{GrunePannek2017MPC}.
In contrast, MPPI typically handles the constraints through penalties in the cost.
Thus, its guarantees depend on the sampling method.
Regarding computational efficiency, RL offers an alternative to optimization-based approaches.
An agent can be trained offline using the model, allowing for fast evaluation on the target machine \cite{egli2021rlexcavatorclassic,andersson2021rlforestrycrane,spinelli2024rlhydraulicmaterialhandling}.
Agents trained with RL have shown high performance: in simulation, on a weakly instrumented excavator, such an agent outperformed both proportional-integral-derivative (PID) and MPPI \cite{roullier2024excavatorcontrolcomparison}.
Hierarchical architectures that combine RL and MPC can take different forms.
For instance, an RL agent can either refine control inputs generated by an MPC \cite{sun2024mpcdrlcommandcorrection} or compensate for model inaccuracies and reject disturbances \cite{zhang2025mpcdrlmodelcorrection}.

Beyond performance, these machines must operate safely.
First, the transfer to the real machine of a policy trained in simulation remains a challenge \cite{aoshima2024sim2realwheelloader}.
Second, RL policies do not provide formal safety guarantees \cite{osinenko2022rlformalguarentee}.
An RL policy thus needs an additional mechanism to operate safely.

Safe RL methods aim to ensure constraint satisfaction during policy execution \cite{wabersisch2023surveysaferl}.
Constrained RL adds the constraints to the training objective.
However, the constraints are then only satisfied in expectation once the training has converged.
Safety filters \cite{wabersisch2018psf} instead correct the actions of the policy online.
Control barrier functions (CBF) ensure that the trajectory of the system remains within a safe set.
These functions can be hard to find and some methods learn them from data \cite{guerrier2024learningcontrolbarrierfunctions}.
Moreover, a CBF is built for a given safe set so that when the constraints change, a new function must be found.
Hamilton--Jacobi (HJ) reachability is another approach \cite{wabersisch2023surveysaferl,borquez2024hjr}.
However, their cost grows exponentially with the state dimension, a problem for implementation on real systems.

Predictive safety filters are a third type of safety filter \cite{wabersisch2018psf,wabersisch2023surveysaferl}.
A PSF solves an MPC problem around the input proposed by a nominal policy, for instance learned with RL \cite{tearle2021predictive}.
It also computes a backup trajectory that satisfies the constraints over a finite horizon.
If the proposed input is safe, it is applied to the system.
Otherwise, it is minimally modified to keep the system within a safe set.
The backup trajectory ensures that the applied input is safe at every time step, with respect to the prediction model.
The constraints enter the optimization problem solved at each time step.
Unlike CBF and HJ reachability based-methods, a PSF handles constraints that change online.
The single-step PSF only corrects the input at the next time step, potentially leading to myopic, overly conservative or oscillatory actions.
Multi-step PSFs correct the inputs over the whole horizon \cite{bejarano2023multistepPSF}, reducing chattering and large corrections.
To do so, the nominal policy is rolled out on the prediction model, which provides predictions of both states and inputs.
However, both single-step and multi-step PSFs only minimize the deviation from the inputs proposed by the policy \cite{wabersisch2018psf,bejarano2023multistepPSF}, and the predicted states are not used to penalize deviations from the intended trajectory.
The filtered trajectory may thus drift away from it after an intervention.

This article considers a weakly-instrumented Caterpillar 966 wheeled loader, that only carries a GPS, a steering angle sensor, and an IMU.
Internal variables such as the engaged gear and the hydraulic pressures are thus not measured, which prevents the use of an accurate analytic model of the machine.
The objective is to track a reference trajectory while satisfying constraints on both the front and rear bodies of the machine.
To this end, we propose a two-levels architecture.
At the low level, a proportional controller tracks the steering command while an RL agent tracks the speed command.
This latter has been trained offline thanks to a learned loader model too costly to embed in an online optimization.
As it absorbs the complexity of the dynamics, the resulting closed loop behaves, as seen by the high level, like two second-order systems described by an inexpensive analytic model.
At the high level, since an RL policy alone provides no safety guarantee, a PSF enforces the constraints, predicting the low-level closed loop with the analytic model whose error is covered by a conservative margin.
This model could also be used in an MPC to track the reference, but the margin would degrade the precision.
The nominal policy is trained with RL using a full simulator embedding the learned model without safety consideration.
The PSF then rolls out this policy on the analytic model to build a nominal plan, the cost penalizing any deviation from it.
Unlike in \cite{sun2024mpcdrlcommandcorrection,zhang2025mpcdrlmodelcorrection}, where RL learns a correction of the MPC control or of its prediction model, the RL agent is the low-level controller and the PSF predicts the closed loop it forms with the machine.
The contributions of this article are the following:
\begin{enumerate}
    \item A hierarchical architecture for weakly instrumented machines, in which a PSF uses an analytic model of the low-level closed loop rather than a learned model to predict the evolution of the machine and roll out the nominal policy.
    \item A low-level RL speed controller for the navigation of a wheeled loader, trained only on a learned model and deployed on the real machine without fine-tuning.
        This is an evolution from \cite{roullier2024excavatorcontrolcomparison} where the low-level controller was evaluated in simulation only.
    \item An enhanced PSF for articulated machines, whose cost penalizes the deviation of both bodies from the nominal plan over the whole horizon, bringing the machine back onto the trajectory
    \item An experimental validation of the whole architecture on the real machine in construction-site conditions.
\end{enumerate}
The rest of the article is organized as follows.
Section~\ref{sec:problem} states the problem and presents the simulation environment.
Section~\ref{sec:lowlevel_rl} designs the low-level controller and evaluates it on the machine.
Section~\ref{sec:highlevelnominal} presents the nominal policy used by PSF controllers introduced in Section~\ref{sec:highlevel_controller}.
Section~\ref{sec:results} presents the simulated and experimental tracking results.
Section~\ref{sec:conclusion} concludes the article.

\section{Problem formulation and prerequisites}
    \label{sec:problem}

In this section, we present the model of the machine, the problem addressed in this article, and the environment in which the controllers are developed.

    \subsection{System model and problem statement}
        \label{sec:problem_statement}

We describe the machine by the nonlinear time-invariant discrete-time dynamics
\begin{equation}
    x_{k+1} = f(x_k, u_k, \zeta_k),
    \label{eq:true_dynamics}
\end{equation}
where $x_k \in \mathbb{R}^9$ and $u_k \in \mathcal{U} \subset \mathbb{R}^4$ are respectively the state and input at time $k$.
The vector $\zeta_k$ gathers unmeasured parameters detailed below.
The map $f$ represents the true dynamics of the machine nonlinear and unknown.
The controllers are therefore designed on learned models of $f$, presented in Section~\ref{sec:simulator}.
We keep the notation $f$ for the machine and use distinct notations for its models.

\begin{figure}[!tb]
    \centering
    \includegraphics[width = 0.4\textwidth]{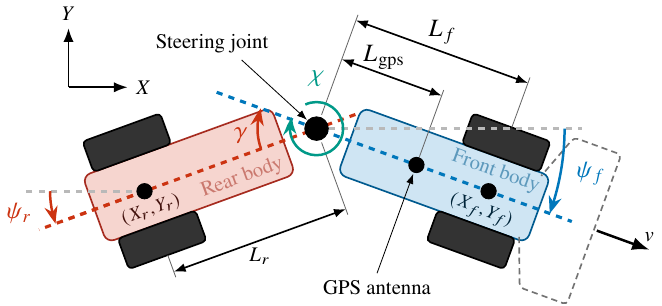}
    \caption{Geometric diagram of the wheeled loader.}
    \label{wheeled_loader_diagram}
\end{figure}

The input $u_k$ represents four physical commands: gear direction (Forward, Neutral or Reverse), engine speed demand, braking effort, and steering direction.
The geometric diagram of the machine is given in Fig.~\ref{wheeled_loader_diagram}.
The state vector is $x_k = \begin{bmatrix}p_k^{\top} & d_k^{\top}\end{bmatrix}^{\top}$.
Its two parts are the following.
\begin{itemize}
    \item $p_k = \begin{bmatrix}X_{\mathrm{f},k} & Y_{\mathrm{f},k} & \psi_{\mathrm{f},k} & X_{\mathrm{r},k} & Y_{\mathrm{r},k} & \psi_{\mathrm{r},k}\end{bmatrix}^{\top}$ is the \emph{kinematic} part, gathering the planar poses of the two bodies in the world frame.
        The front pose $\begin{bmatrix}X_{\mathrm{f}} & Y_{\mathrm{f}} & \psi_{\mathrm{f}}\end{bmatrix}^{\top}$ is taken at the GPS antenna, located on the front body at a distance $L_{\mathrm{gps}}$ from the steering joint.
        The rear pose $\begin{bmatrix}X_{\mathrm{r}} & Y_{\mathrm{r}} & \psi_{\mathrm{r}}\end{bmatrix}^{\top}$ is taken at the rear axle.
    \item $d_k = \begin{bmatrix}v_k & \gamma_k & \chi_k\end{bmatrix}^{\top}$ is the \emph{dynamic} part, gathering the longitudinal speed $v$, the steering angle $\gamma = \psi_{\mathrm{f}} - \psi_{\mathrm{r}}$ and its velocity $\chi = \dot{\gamma}$.
\end{itemize}
We keep both poses in the state because our constraints apply to both bodies.
The state is obtained from the GPS and the IMU.
The rear pose is reconstructed from the front pose and $\gamma$.
The state vector is a modeling choice and does not contain all the internal variables of the machine.
The inacessible variables such as the engaged gear, the hydraulic pressures and even the soil conditions are gathered in $\zeta_k$.
None of them are measured.
As a consequence, $x_k$ and $u_k$ alone do not determine $x_{k+1}$.
The learned model of Section~\ref{sec:simulator} compensates for $\zeta_k$ with a history of past measurements and commands.
The machine must track a finite time-indexed reference trajectory $\{x^{\mathrm{ref}}_k\}_{k=0}^{N_{\mathrm{ref}}}$ such that ${x^{\mathrm{ref}}_k}^{\top} = \begin{bmatrix}{p^{\mathrm{ref}}_k}^{\top} & v^{\mathrm{ref}}_k\end{bmatrix}$, where $p^{\mathrm{ref}}_k = \begin{bmatrix}X^{\mathrm{ref}}_{\mathrm{f},k} & Y^{\mathrm{ref}}_{\mathrm{f},k} & \psi^{\mathrm{ref}}_{\mathrm{f},k} & X^{\mathrm{ref}}_{\mathrm{r},k} & Y^{\mathrm{ref}}_{\mathrm{r},k} & \psi^{\mathrm{ref}}_{\mathrm{r},k}\end{bmatrix}^{\top}$ gathers the reference poses of both bodies and $v^{\mathrm{ref}}_k$ is the reference speed.
The state is constrained to a set $\mathcal{X} \subset \mathbb{R}^9$ described in Section~\ref{sec:highlevel_controller} to remain in operation area while avoiding obstacles.
The problem is then to compute inputs $u_k \in \mathcal{U}$ such that both bodies track the reference trajectory with $x_k \in \mathcal{X}$ for all $k$.

    \subsection{Control architecture}
        \label{sec:architecture}

\begin{figure*}[!tb]
    \centering
    \includegraphics[width = \textwidth]{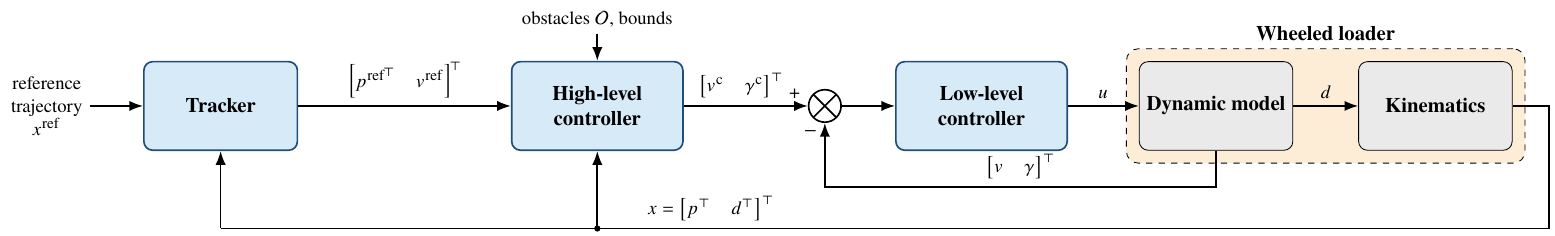}
    \caption{Control architecture.}
    \label{control_architecture_diagram}
\end{figure*}

To address this problem, we use the hierarchical architecture of Fig.~\ref{control_architecture_diagram}.
At each time step $k$, a tracker provides the PSF of the high-level controller with the slice of the reference trajectory $x^{\mathrm{ref}}$ that starts at $x^{\mathrm{ref}}_k$ and extends over a given horizon.
Since the reference trajectory is indexed by time, this slice is selected from the time elapsed since the start of the experiment.
However, if the machine falls too far behind, the tracker holds the reference and waits for the machine, so that the gap between the machine and the reference never exceeds a maximum leash (parameter given by the user).
This controller provides a high-level speed and steering command $r_k = \begin{bmatrix} v^{\mathrm{c}}_k & \gamma^{\mathrm{c}}_k \end{bmatrix}^{\top}$ to the low-level controller.
This command is calculated to keep the machine safe while staying as close as possible to $x^{\mathrm{ref}}_k$.
The low-level controller then computes the input $u_k$ of the machine.
The two levels run at different rates.
The low-level controller runs at \qty{100}{\hertz} and the high-level controller runs at \qty{4}{\hertz}, each command $r_k$ being held constant over the 25 low-level steps.
For simplicity, $k$ denotes the time step of the level considered: \qty{0.01}{\second} for low-level and \qty{0.25}{\second} for high-level.
The wheeled loader block of Fig.~\ref{control_architecture_diagram} is either the simulator or the machine, depending on the context.

    \subsection{Simulation environment}
        \label{sec:simulator}

The numerical simulator approximating the dynamics of \eqref{eq:true_dynamics} combines two parts :
The first one is a neural network $f_\theta$ giving the dynamic state $d$ from the commands $u_k$.
The second part is the analytic kinematic model $f^{\mathrm{p}}$ reconstructing the poses $p$ of both bodies from $d$.
Together, $f_\theta$ and $f^{\mathrm{p}}$ predict the full state $x_{k+1}$.

        \subsubsection{Dynamic model}

The dynamic state is propagated by the learned map $d_{k+1} = d_k + f_\theta\left(z_k\right)$, where the network predicts the increment of the dynamic state.
Its input is the stacked history $z_k = \left\{ \left(u_{k-j}, d_{k-j}\right) \right\}_{j \in \mathcal{J}}$, where $\mathcal{J}$ is the set of $N_{\mathrm{s}}$ past time steps selected by the stack.
Analytic wheeled loader models \cite{zhang2020analyticmodelwheelloader,kumar2024operatormodelwheeledloader} rely on physical parameters that drift with wear and soil conditions, and on sensors that our machine lacks.
We therefore learn $f_\theta$ from data recorded on the machine.
Following previous studies \cite{hoffmann2025lstmexcavator}, we choose to represent $f_\theta$ as a stacked-input LSTM.
The model is trained on a dataset collected from the machine itself at \qty{100}{\hertz}, which includes a wide range of operating conditions and ground types to ensure good generalization capabilities.
In addition to that, the recordings were carried out over an extended period of time, resulting in experiments performed on significantly different soil conditions.
This temporal spacing further increases the diversity of the dataset and improves its representativeness across varying environmental conditions.
Once the data is collected and processed, the LSTM model was trained with a suitable loss function capturing the prediction error of the state variables.
Hyperparameter tuning is performed with Bayesian optimization to find the optimal architecture and training parameters \cite{victoria2021bayesianoptimizationhp}.
The values of the network hyperparameters are given in Table~\ref{tab:hyperparams}.
The mean absolute error (MAE), the root mean square error (RMSE), the 95th percentile (P95), and the coefficient of determination ($R^2$) are considered to assess the model's performance on two tests:

\begin{table}[t]
    \centering
    \caption{LSTM model and training hyperparameters}
    \label{tab:hyperparams}
    \footnotesize
    \begin{tabular}{@{}ll ll@{}}
        \toprule
        \textbf{Hyperparameter} & \textbf{Value} &
        \textbf{Hyperparameter} & \textbf{Value} \\
        \midrule
        Model type                  & LSTM                & Hidden layers     & 3 \\
        Hidden layer width          & 500                 & Dropout rate      & 0.25 \\
        Time step duration          & \qty{0.01}{\second} & Batch size        & 64 \\
        Stack size $N_{\mathrm{s}}$ & 20                  & Stack step        & 10 \\
        Window size                 & 500                 & Stack growth rate & Linear \\
        Learning rate               & 0.001               & No-loss period    & 15.0 \\
        \bottomrule
    \end{tabular}
\end{table}

            \paragraph*{Open-loop test}

We apply inputs provided by the low-level controller of Section~\ref{sec:lowlevel_rl}, to both the machine and the model.
The errors are reported in Table~\ref{tab:model_results}.
The steering angle and velocity remain close to the measurements.
The speed drifts further as in open loop, nothing corrects the prediction and any small error gets integrated.
Despite this drift, the errors remain moderate over the whole run.
The model is thus accurate enough to develop the controllers of the next sections.

\begin{figure}[!tb]
    \centering
    \includegraphics[width = 0.9\columnwidth]{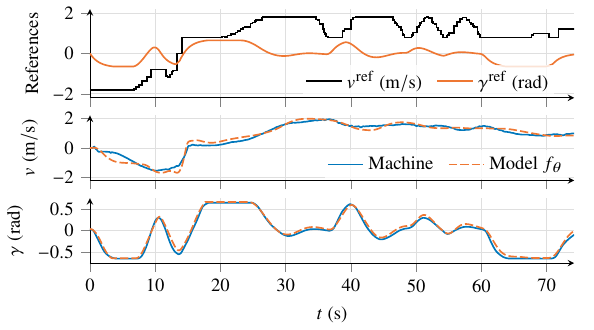}
    \caption{Closed-loop validation of $f_\theta$ against the machine.}
    \label{lstm_drl_injected_closed_loop}
\end{figure}

\begin{table}[t]
    \centering
    \caption{Prediction errors on the machine data}
    \label{tab:model_results}
    \footnotesize
    \begin{tabular}{@{}l l l l l@{}}
        \toprule
        \textbf{Test} & \textbf{Metric} & \textbf{Speed}             & \textbf{Steer. angle} & \textbf{Steer. velocity} \\
                      &                 & (\unit{\metre\per\second}) & (\unit{\radian})      & (\unit{\radian\per\second}) \\
        \midrule
        Open loop   & RMSE  & 0.404 & 0.045 & 0.031 \\
                    & MAE   & 0.347 & 0.035 & 0.024 \\
                    & P95   & 0.702 & 0.085 & 0.061 \\
                    & $R^2$ & 0.740 & 0.985 & 0.955 \\
        \midrule
        Closed loop & RMSE  & 0.190 & 0.051 & 0.036 \\
                    & MAE   & 0.153 & 0.042 & 0.029 \\
                    & P95   & 0.378 & 0.091 & 0.069 \\
                    & $R^2$ & 0.961 & 0.984 & 0.939 \\
        \bottomrule
    \end{tabular}
\end{table}

            \paragraph*{Closed-loop test}

We use a proportional speed controller with a first-order memory on the previous command and a proportional steering controller to drive both the model and the machine in closed loop.
Both track the same commands, as illustrated in Fig.~\ref{lstm_drl_injected_closed_loop}, and the performances are reported in Table~\ref{tab:model_results}.
The speed RMSE drops to \qty{0.19}{\meter\per\second} and its $R^2$ rises to \num{0.96}.
The controller corrects the speed drift observed in open loop.
The steering angle and velocity errors remain close to those in the previous test.

        \subsubsection{Kinematics}

The poses follow the articulated-vehicle kinematic model \cite{nayl2012wheelloadermodel,Penttila2025wheelloadermpcrl} $\dot{p} = f^{\mathrm{p}}(p, d)$,
\begin{subequations}
    \label{eq:kinematic_model}
    \allowdisplaybreaks
    \begin{align}
        \label{eq:yaw_front}
        \dot{\psi}_{\mathrm{f}} & = \left(v \sin\gamma + L_r \chi\right)/\left(L_f \cos\gamma + L_r\right),\\
        \label{eq:x_front}
        \dot{X}_{\mathrm{f}}    & = v \cos\psi_{\mathrm{f}} - (L_{\mathrm{gps}} - L_f) \dot{\psi}_{\mathrm{f}} \sin\psi_{\mathrm{f}},\\
        \label{eq:y_front}
        \dot{Y}_{\mathrm{f}}    & = v \sin\psi_{\mathrm{f}} + (L_{\mathrm{gps}} - L_f) \dot{\psi}_{\mathrm{f}} \cos\psi_{\mathrm{f}},\\
        \label{eq:yaw_rear}
        \dot{\psi}_{\mathrm{r}} & = \dot{\psi}_{\mathrm{f}} - \chi,\\
        \label{eq:x_rear}
        \dot{X}_{\mathrm{r}}    & = \dot{X}_{\mathrm{f}} + L_{\mathrm{gps}} \dot{\psi}_{\mathrm{f}} \sin\psi_{\mathrm{f}} + L_r \dot{\psi}_{\mathrm{r}} \sin(\psi_{\mathrm{f}} - \gamma),\\
        \label{eq:y_rear}
        \dot{Y}_{\mathrm{r}}    & = \dot{Y}_{\mathrm{f}} - L_{\mathrm{gps}} \dot{\psi}_{\mathrm{f}} \cos\psi_{\mathrm{f}} - L_r \dot{\psi}_{\mathrm{r}} \cos(\psi_{\mathrm{f}} - \gamma),
    \end{align}
\end{subequations}
where $L_f$ and $L_r$ are the front and rear wheelbases, as illustrated in Fig.~\ref{wheeled_loader_diagram}.
The distance $L_{\mathrm{gps}}$ from the joint to the GPS antenna, defined in Section~\ref{sec:problem_statement}, is identified on GPS data recorded on the machine.
In the simulator, $f^{\mathrm{p}}$ is integrated over each time step with $v$, $\gamma$ and $\chi$ predicted by $f_\theta$.
Coupled with the low-level controller, the simulator takes the high-level commands $r_k$ as inputs.

\section{Low-level controller design with reinforcement learning}
    \label{sec:lowlevel_rl}

In this section, we design the low-level controller tracking the speed and steering commands sent by the high-level controller.
The controller must track the speed command precisely with smooth inputs, as abrupt ones can damage the machine's transmission.
Unlike an online optimization embedding $f_\theta$, an LSTM RL agent trained offline is fast to evaluate on the machine \cite{egli2021rlexcavatorclassic,andersson2021rlforestrycrane}.
We therefore use one for the speed loop, while the steering loop keeps a proportional controller.
Formally, the low-level controller is the map
\begin{equation}
    u_k = \pi_{\mathrm{ll}}\left(r_k, d_k, m_k\right)
    \label{eq:map_lowlevel}
\end{equation}
where $m_k$ is a memory term gathering the recurrent's policy internal state and last two actions, updated at each time step.

    \subsection{Agent architecture}

The policy is an LSTM, whose architecture is given in Table~\ref{tab:arch_ll}.
It maps the observation to an action $u^{\mathrm{ped}}_k \in [-1,1]$.
Its sign selects the gear direction, forward when positive and reverse when negative.
Its magnitude is the engine speed demand.
The braking effort is binary.
It is set to 0 when $|u^{\mathrm{ped}}_k| > 0.05$ and to 1 otherwise.
The machine responds with delays, and part of its state is not measured (such as the selected gear ratio).
The internal state of the LSTM accounts for the past behavior of the machine.
It replaces an explicit history in the observation, which remains compact.
We also expect smoother commands than with a memoryless policy, such as the multilayer perceptron of \cite{egli2021rlexcavatorclassic}.
The agent is only in charge of the longitudinal control: the steering direction is given by a proportional map on the target steering angle $\gamma^{\mathrm{c}}_k$.
A low-pass filter smooths this command to avoid damaging the steering joint.
At each step $k$, the agent receives four observations: the longitudinal speed error $e_k = v_k - v^{\mathrm{c}}_k$, the current steering angle $\gamma_k$ and velocity $\chi_k$, as well as the last action increment $s_a \lvert \Delta u^{\mathrm{ped}}_{k-1} \rvert$.
Here $\Delta u^{\mathrm{ped}}_k = u^{\mathrm{ped}}_k - u^{\mathrm{ped}}_{k-1}$ and $s_a$ is a scaling factor.
Providing $\gamma_k$ and $\chi_k$ allows the agent to adapt the speed to hard steering maneuvers such as sharp turns.
In addition to that, the action increment helps to smooth the control actions over time.

    \subsection{Agent training}

The agent is trained with proximal policy optimization (PPO) \cite{schulman2017ppo}, the LSTM model of Section~\ref{sec:simulator} being used as the training environment.
The training hyperparameters are given in Table~\ref{tab:train_ll}.
The reward reads
\begin{align}
    \label{eq:ll_reward}
    \mathcal{R}^{\mathrm{ll}}_k = &-w_v\left(1 - \exp(-|e_k|/\sigma_v)\right)
        - w_{\mathrm{s}} |u^{\mathrm{ped}}_k| \mathds{1}_{\left\{\substack{\operatorname{sgn} u^{\mathrm{ped}}_k = \operatorname{sgn} e_k\\ |e_k| \geq \epsilon_e}\right\}} \notag\displaybreak[0]\\
    &- w_{\mathrm{b}} |v_k| \mathds{1}_{\{|v^{\mathrm{c}}_k| < v_{\mathrm{stop}}\}}
        - w_{\Delta} |\Delta u^{\mathrm{ped}}_k| \notag\displaybreak[0]\\
    &- w_{\mathrm{ped}} |u^{\mathrm{ped}}_k| - w_{\mathrm{o}} \Omega_k,
\end{align}
where $\mathds{1}_{S}$ is the indicator function of the set $S$.
The first term penalizes the speed tracking error; its bounded exponential shape prevents large errors from overwhelming the other terms and avoids the bang-bang behavior induced by a linear penalty.
The second term penalizes the engine speed demand when the gear direction drives the speed away from its command, outside a dead zone $\epsilon_e$.
The third one penalizes the residual speed when a stop is requested.
The last three terms promote smooth commands.
At each step, $w_{\mathrm{ped}}$ penalizes the engine speed demand and $w_\Delta$ its variation.
$\Omega_k$ captures slower limit cycles that a one-step variation does not see.
It is the mean turnaround amplitude of the engine speed demand over a sliding window of $T_{\mathrm{o}}$ seconds,
\begin{equation}
    \Omega_k = \frac{1}{|\mathcal{D}_k| - 1}\sum_{i} \min\left(|\Delta u^{\mathrm{ped}}_i|, |\Delta u^{\mathrm{ped}}_{i-1}|\right) \mathds{1}_{\{\Delta u^{\mathrm{ped}}_i \Delta u^{\mathrm{ped}}_{i-1} < 0\}},
    \label{eq:ll_oscillation}
\end{equation}
where $\mathcal{D}_k = \{\Delta u^{\mathrm{ped}}_i\}$ is the set of nonzero increments of the engine speed demand within the window.
A monotone ramp costs nothing, only back-and-forth motions are penalized.

The episodes correspond to diverse speed and steering angle profiles.
These profiles differ in terms of speed and steering angle amplitudes, as well as the rate of change of these variables.
Those parameters are gradually increased during the training process to progressively challenge the agent and improve its generalization capabilities.
This process, called curriculum learning, has already been tested in several contexts \cite{wang2022surveycurriculumlearning}.
The episodes have the same duration and are initialized with the same initial state, i.e., the vehicle starts from rest and has a zero steering angle.

\begin{table}[t]
    \centering
    \caption{Low-level RL agent network architecture}
    \label{tab:arch_ll}
    \footnotesize
    \begin{tabular}{@{}l l l l@{}}
        \toprule
        \textbf{Parameter} & \textbf{Value} & \textbf{Parameter} & \textbf{Value} \\
        \midrule
        Model type    & LSTM & Activation function & tanh \\
        Hidden layers & 2    & Hidden size         & 128  \\
        \bottomrule
    \end{tabular}
\end{table}

\begin{table}[t]
    \centering
    \caption{Low-level RL agent training hyperparameters}
    \label{tab:train_ll}
    \footnotesize
    \begin{tabular}{@{}ll ll@{}}
        \toprule
        \textbf{Hyperparameter} & \textbf{Value} &
        \textbf{Hyperparameter} & \textbf{Value} \\
        \midrule
        Steps per update   & \num{4096}   & Discount factor                  & \num{0.98} \\
        Batch size         & \num{64}     & Generalized advantage estimation & \num{0.95} \\
        Learning rate      & \num{0.0003} & Clipping parameter               & \num{0.2}  \\
        Epochs per update  & \num{10}     & Entropy coef.                    & \num{0.03} \\
        Max. gradient norm & \num{0.5}    & Value function coef.             & \num{0.5}  \\
        \bottomrule
    \end{tabular}
\end{table}

    \subsection{Experimental evaluation on the real machine}

The agent is deployed on the real wheeled loader without fine-tuning.
Since the agent never interacted with the machine during training, this zero-shot transfer also indirectly validates $f_\theta$.
We compare the agent's policy with PID controllers, on the same machine.
The comparison is made in two tests.

            \paragraph*{Sinusoidal tracking}

\begin{figure}[tb]
    \centering
    \includegraphics[width = .9\columnwidth]{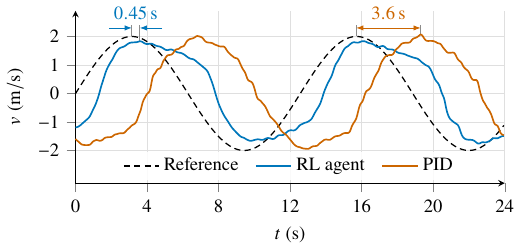}%
    \caption{Sinusoidal speed command on the real wheeled loader.}
    \label{ll_sinusoid_pid_vs_ppo}
\end{figure}

Both low-level speed controllers receive the same speed command profile, a sinusoid of amplitude \qty{2}{\meter\per\second} and period \qty{12.6}{\second}.
The speed command changes sign, so the machine alternates between forward and reverse motions.
Both runs are compared in Fig.~\ref{ll_sinusoid_pid_vs_ppo} over the second and third periods of the command, after the start-up transient.
The RL agent follows the command with a delay of about \qty{1.3}{\second}, against \qty{3.8}{\second} for the PID controller.
Its speed RMSE is \qty{0.91}{\meter\per\second}, against \qty{2.35}{\meter\per\second} for the PID controller, most of the error coming from the delay.

            \paragraph*{Production cycle}

\begin{figure}[tb]
    \centering
    \includegraphics[width = .9\columnwidth]{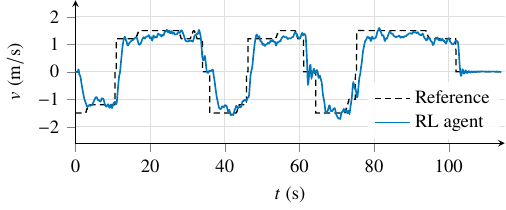}
    \caption{Tracking speed of the RL agent during a production cycle.}
    \label{ll_pid_vs_rl}
\end{figure}

The machine performs a production cycle under industrial operating conditions.
All runs share the same high-level Stanley controller \cite{hoffmann2007stanley} and reference trajectory.
The RL agent is compared with three PID controllers.
PID~1 is tuned to give the smoothest possible commands.
PID~2 is tuned directly on the machine with the Ziegler-Nichols method and PID~3 is tuned in the simulator of Section~\ref{sec:simulator} to give heuristically the best speed tracking.
Fig.~\ref{ll_pid_vs_rl} shows the speed tracking of the RL agent and Table~\ref{tab:ll_performances} gives the metrics of all runs.
Besides the RMSE, MAE and P95 of the speed error, it reports the engine speed demand variation defined as the sum of the absolute increments of the engine speed demand, divided by the duration of the run.
The RL agent has the lowest RMSE, MAE and P95.
PID~2 and PID~3 come close, but with a much larger engine speed demand variation.
PID~1 has smoother commands than the RL agent, but a much larger tracking error.
Because of the recurrent nature of its LSTM policy, the RL agent is the only controller that is both accurate and relatively smooth.
That is why it will now be used as a low-level controller for speed tracking.

\begin{table}[tb]
    \centering
    \caption{Speed tracking metrics over a production cycle}
    \label{tab:ll_performances}
    \footnotesize
    \begin{tabular}{@{}l c c c c@{}}
        \toprule
        \textbf{Controller} & \textbf{RMSE}                & \textbf{MAE}               & \textbf{P95}               & \textbf{Demand var.} \\
                            & (\unit{\meter\per\second})   & (\unit{\meter\per\second}) & (\unit{\meter\per\second}) & (\unit{\hertz}) \\
        \midrule
        RL agent & \textbf{0.511} & \textbf{0.289} & \textbf{1.367} & \num{0.111} \\
        \midrule
        PID 1 & \num{0.709} & \num{0.512} & \num{1.447} & \textbf{0.035} \\
        PID 2 & \num{0.569} & \num{0.300} & \num{1.433} & \num{0.258} \\
        PID 3 & \num{0.592} & \num{0.374} & \num{1.439} & \num{0.710} \\
        \bottomrule
    \end{tabular}
\end{table}

\section{High-level nominal policy design}
    \label{sec:highlevelnominal}

The high-level controller computes the high-level speed and steering commands $r_k = \begin{bmatrix} v^{\mathrm{c}}_k & \gamma^{\mathrm{c}}_k \end{bmatrix}^{\top}$ sent to the low-level controller $\pi_{\mathrm{ll}}$ of \eqref{eq:map_lowlevel}.
Its core is the enhanced PSF of Section~\ref{sec:highlevel_controller}, which enforces the constraints and brings both bodies back onto the plan of a nominal policy after an intervention.
This section designs this nominal policy, which only has to track the reference trajectory, safety being entirely handled by the filter.
Since the filter queries it at each step of its prediction horizon $H$, the nominal policy must be cheap to evaluate, which motivates the use of an RL agent.
Section~\ref{sec:nominal_elaboration} details its elaboration, and Section~\ref{sec:nominal_machine} evaluates it on the loader.

    \subsection{Nominal controller elaboration}
        \label{sec:nominal_elaboration}

\begin{figure}[tb]
    \centering
    \includegraphics[width = .7\columnwidth]{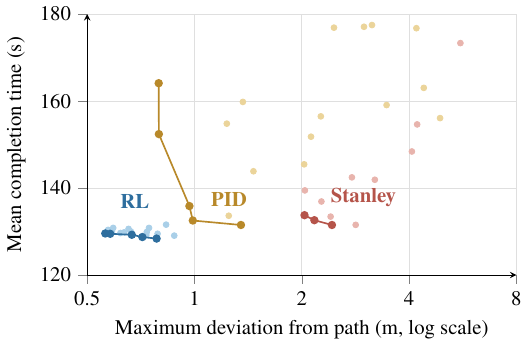}
    \caption{Pareto comparison of the agents on industrial criteria.}
    \label{fig:pareto_comparison}
\end{figure}

The agent is trained with PPO \cite{schulman2017ppo} in the simulator of Section~\ref{sec:simulator}, driven by the low-level controller of Section~\ref{sec:lowlevel_rl}.
The architecture of the agent is given in Table~\ref{tab:arch_hl} and its training parameters are given in Table~\ref{tab:train_hl}.
At each step $k$, the agent compares the current state of the machine with the $H$ next points of the reference trajectory.
For each body $b \in \{\mathrm{f},\mathrm{r}\}$ and each lookahead step $i = 0, \dots, H - 1$, let $\Delta X_{b,k+i} = X_{b,k} - X^{\mathrm{ref}}_{b,k+i}$ and $\Delta Y_{b,k+i} = Y_{b,k} - Y^{\mathrm{ref}}_{b,k+i}$ be the position offsets in the world frame.
The tracking errors are expressed in the body's frame as
\begin{subequations}
    \label{eq:hl_block_errors}
    \begin{align}
        \label{eq:hl_elon}
        e^{\mathrm{lon}}_{b,k+i} & = \phantom{-}\cos\psi_{b,k} \Delta X_{b,k+i} + \sin\psi_{b,k} \Delta Y_{b,k+i},\\
        \label{eq:hl_elat}
        e^{\mathrm{lat}}_{b,k+i} & = -\sin\psi_{b,k} \Delta X_{b,k+i} + \cos\psi_{b,k} \Delta Y_{b,k+i},\\
        \label{eq:hl_epsi}
        e^{\psi}_{b,k+i}         & = \kappa_{\deg} \mathrm{wrap}\left(\psi_{b,k} - \psi^{\mathrm{ref}}_{b,k+i}\right)
    \end{align}
\end{subequations}
with $\mathrm{wrap}(\theta) = \left((\theta+\pi) \bmod 2\pi\right) - \pi$ and $\kappa_{\deg} = 180/\pi$.
This separates a lag along the path ($e^{\mathrm{lon}}$) from a lateral deviation ($e^{\mathrm{lat}}$), which can then be weighted independently, and the wrapping prevents two headings differing by $2\pi$, i.e., the same physical state, from being penalized.
For each $i$, the observation of the agent gathers the errors $e^{\mathrm{lon}}_{\mathrm{f},k+i}$, $e^{\mathrm{lat}}_{\mathrm{f},k+i}$ and $e^{\psi}_{\mathrm{f},k+i}$ of the front body, the heading error $e^{\psi}_{\mathrm{r},k+i}$ of the rear body, whose heading is computed as $\psi_{\mathrm{r},k} = \psi_{\mathrm{f},k} - \gamma_k$, and the speed error $e^{v}_{k+i} = v_k - v^{\mathrm{ref}}_{k+i}$.
It also contains the steering angle $\gamma_k$, the steering velocity $\chi_k$, and the previous commands $v^{\mathrm{c}}_{k-1}$ and $\gamma^{\mathrm{c}}_{k-1}$ given to the low-level controller (set to 0 at reset), which are kept constant over the lookahead steps.
The observation therefore has $9H$ components, where we chose $H = 15$, i.e., a preview of \qty{3.75}{\second}.
Unlike the observation, the reward only uses the current reference point, i.e., $i = 0$, so that the progression of the reference along the path is not counted as a tracking error.
It penalizes the errors of both bodies, with the heading errors converted to degrees by the factor $\kappa_{\deg}$.
Let $c^{\mathrm{ref}}_k$ be the signed curvature of the reference path at step $k$.
The reference steering angle $\gamma^{\mathrm{ref}}_k$ is the steering angle that holds this curvature at steady state, i.e., with $\chi = 0$ in \eqref{eq:yaw_front}.
It is the solution of $c^{\mathrm{ref}}_k = \sin\gamma^{\mathrm{ref}}_k / (L_f \cos\gamma^{\mathrm{ref}}_k + L_r)$, clipped to the steering range.
This relation holds for the front axle, but we apply it to the path of the GPS antenna, since both curvatures only differ by a second-order term in $(L_{\mathrm{gps}} - L_f)$.
Let $\Delta v^{\mathrm{c}}_k = v^{\mathrm{c}}_k - v^{\mathrm{c}}_{k-1}$ and $\Delta \gamma^{\mathrm{c}}_k = \gamma^{\mathrm{c}}_k - \gamma^{\mathrm{c}}_{k-1}$ be the command increments.
The reward then reads as
\begin{subequations}
    \label{eq:reward_hl_total}
    \begin{align}
        \mathcal{R}^{\mathrm{hl}}_k & = -\sum_{b \in \{\mathrm{f},\mathrm{r}\}} \left[ \alpha_{\mathrm{lon}} \left(e^{\mathrm{lon}}_{b,k}\right)^2 + \alpha_{\mathrm{lat}} \left(e^{\mathrm{lat}}_{b,k}\right)^2 + \alpha_{\psi} \left(e^{\psi}_{b,k}\right)^2 \right] \notag\displaybreak[0]\\
        & \phantom{=\;} - \alpha_v \left(e^{v}_{k}\right)^2 - \beta_\gamma |\gamma^{\mathrm{c}}_k - \gamma^{\mathrm{ref}}_k| - \beta_v |v^{\mathrm{c}}_k - v^{\mathrm{ref}}_k| \notag\displaybreak[0]\\
        \label{eq:reward_hl}
        & \phantom{=\;} - \lambda_\gamma \rho^{\gamma}_k - \lambda_v \rho^{v}_k - \eta_{\mathrm{w}} w_k - \eta_{\mathrm{s}} s_k - \nu_\gamma o^{\gamma}_k - \nu_v o^{v}_k,\\
        \label{eq:reward_hl_rate}
        \rho^{\gamma}_k &= \max(0, |\Delta\gamma^{\mathrm{c}}_k| - \epsilon_\gamma), \;
            \rho^{v}_k = \max(0, |\Delta v^{\mathrm{c}}_k| - \epsilon_v),\\
        \label{eq:reward_hl_wrongway}
        w_k &= |v^{\mathrm{c}}_k| \mathds{1}_{\left\{\substack{|v^{\mathrm{ref}}_k| > \epsilon_{\mathrm{w}}\\v^{\mathrm{c}}_k v^{\mathrm{ref}}_k < 0}\right\}},\quad
            s_k = |v^{\mathrm{c}}_k| \mathds{1}_{\left\{\substack{|v_k| > \epsilon_{\mathrm{s}}\\v^{\mathrm{c}}_k v_k < 0}\right\}},\\
        \label{eq:reward_hl_osc}
        \MoveEqLeft[1.5] \sum_{k \in \mathcal{E}^{\gamma}_j} o^{\gamma}_k = \max\left(0, \min\left(L^{\gamma,-}_j, L^{\gamma,+}_j\right) - \epsilon^{\gamma}_{\mathrm{osc}}\right).
    \end{align}
\end{subequations}
The weights $\alpha_{\mathrm{lon}}$, $\alpha_{\mathrm{lat}}$, $\alpha_{\psi}$, $\alpha_v$, $\beta_\gamma$, $\beta_v$, $\lambda_\gamma$, $\lambda_v$, $\eta_{\mathrm{w}}$, $\eta_{\mathrm{s}}$, $\nu_\gamma$ and $\nu_v$ are positive real numbers.
The first line of \eqref{eq:reward_hl} penalizes the pose errors of both bodies, the rear terms making the agent bring the whole machine onto the path, not just the front body.
The second line of \eqref{eq:reward_hl} penalizes the speed error and pulls the commands toward the reference speed and steering angle, i.e., toward an angle of zero on a straight path and the angle required by the curve otherwise.
The last line of \eqref{eq:reward_hl} penalize the increments of the commands beyond the deadbands $\epsilon_\gamma$ and $\epsilon_v$ of \eqref{eq:reward_hl_rate}, as well as three faults of the commands.
The deadbands $\epsilon_\gamma = \qty{0.05}{\radian}$ and $\epsilon_v = \qty{0.1}{\meter\per\second}$ are the command increments allowed per step without penalty.
They let the agent ramp the steering and follow the accelerations of the reference, but penalize abrupt jumps.
The terms $w_k$ and $s_k$, defined in \eqref{eq:reward_hl_wrongway}, penalize a speed command opposite to the reference speed and to the measured speed, respectively: in the latter case, the command reverses the gearbox while the machine is moving, and the speed then responds with a dead time and an overshoot.
Finally, $o^{\gamma}_k$ and $o^{v}_k$ penalize the back-and-forth motions of the commands.
In \eqref{eq:reward_hl_osc}, $\mathcal{E}^{\gamma}_j$ is the set of steps spanned by the $j$-th reversal of the direction of the steering command, and $L^{\gamma,-}_j$ and $L^{\gamma,+}_j$ are the steering travels before and after it.
The minimum of these two travels measures the part of the steering motion that is cancelled by the reversal; this motion is wasted, since it only makes the machine snake around the path.
The allowance $\epsilon^{\gamma}_{\mathrm{osc}}$ leaves small corrections free.
Unlike $\rho^{\gamma}_k$, this term also penalizes slow oscillations.
The term $o^{v}_k$ is defined in the same way on the speed command, with the allowance $\epsilon^{v}_{\mathrm{osc}}$.
A monotone ramp of a command is thus never penalized.
To justify our choice, the RL agent is compared with two classical controllers that are equally inexpensive to evaluate.
The first one consists of two PID controllers, which compute the speed and steering commands from the longitudinal and lateral errors of the front body, computed on reference points located $T_1$ and $T_2$ seconds ahead, respectively.
The second one uses a Stanley controller for the steering, and the same PID law as above for the longitudinal speed.
All the controllers are tested in the same simulator for several settings of their parameters: gains and preview horizons $T_1$ and $T_2$ for the classical controllers, and reward functions and hyperparameters for the RL agents.
Each setting is run on 10 reference trajectories and evaluated on precision and speed, the two main industrial criteria.
Precision is measured by the maximum path deviation, i.e., the largest distance to the closest point of the reference path, and speed by the mean time taken to complete the trajectory.
The resulting Pareto fronts, illustrated in Fig.~\ref{fig:pareto_comparison}, show that the RL-agent-based approach dominates both classical ones.
Therefore, it is used to design the high-level nominal policy.

\begin{table}[t]
    \centering
    \caption{High-level RL agent network architecture}
    \label{tab:arch_hl}
    \footnotesize
    \begin{tabular}{@{}l l l l@{}}
        \toprule
        \textbf{Parameter} & \textbf{Value} & \textbf{Parameter} & \textbf{Value} \\
        \midrule
        Model type    & LSTM & Activation function & tanh \\
        Hidden layers & 1    & Hidden size         & 256  \\
        \bottomrule
    \end{tabular}
\end{table}

\begin{table}[t]
    \centering
    \caption{High-level RL agent training hyperparameters}
    \label{tab:train_hl}
    \footnotesize
    \begin{tabular}{@{}ll ll@{}}
        \toprule
        \textbf{Hyperparameter} & \textbf{Value} &
        \textbf{Hyperparameter} & \textbf{Value} \\
        \midrule
        Steps per update   & \num{2048}   & Discount factor       & \num{0.99} \\
        Batch size         & \num{256}    & Generalized advantage & \multirow{2}{*}{\num{0.95}} \\
        Learning rate      & \num{0.0001} & estimation            & \\
        Epochs per update  & \num{5}      & Clipping parameter    & \num{0.2} \\
        Max. gradient norm & \num{0.5}    & Entropy coef.         & \num{0.01} \\
        Target KL div.     & \num{0.02}   & Value function coef.  & \num{0.5} \\
        \bottomrule
    \end{tabular}
\end{table}

    \subsection{Experimental evaluation of the nominal policy}
        \label{sec:nominal_machine}

\begin{figure}[tb]
    \centering
    \includegraphics[width = .9\columnwidth]{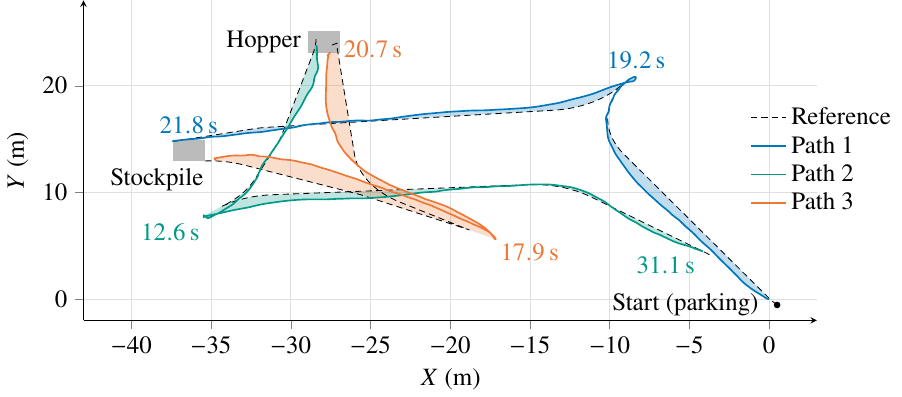}
    \caption{Path tracking with the nominal policy on three paths.}
    \label{fig:drl_machine_trajectory}
\end{figure}

The nominal RL policy is first deployed on the wheeled loader, initially without any filter.
It is tested on several trajectories, as illustrated in Fig.~\ref{fig:drl_machine_trajectory}.
The figure shows the reference path and the path of the front body.
The first path goes from a parking area to a stockpile.
The machine first reverses away from the parking area, then drives forward to the stockpile.
The curvature of this path is continuous.
The second path starts at a hopper.
The machine first reverses away from the hopper, then drives forward back to a parking area.
The third path goes from a stockpile to a hopper.
The machine first reverses away from the stockpile, then drives forward to the hopper.
The metrics, computed on the front position $\begin{bmatrix}X_{\mathrm{f}} & Y_{\mathrm{f}}\end{bmatrix}^{\top}$, of each path are given in Table~\ref{tab:drl_machine_metrics}.
The path deviation is the distance from this point to the closest point of the reference path, and thus only measures how far the machine strays from the path.
The tracking error is the Euclidean distance from this point to the reference point $\begin{bmatrix}X^{\mathrm{ref}}_{\mathrm{f},k} & Y^{\mathrm{ref}}_{\mathrm{f},k}\end{bmatrix}^{\top}$ at the same time, so that it also captures the delay of the machine along the path.
Finally, the completion time is measured from the first reference point until the machine stops at the end of the path.
The first two paths are feasible by the machine, since their initial curvature is continuous with the end of the previous path, or with the parked configuration of the machine for the first path.
On these two paths, the maximum path deviation remains below \qty{1.6}{m}.
On the second path, it is reached at the cusp, i.e., the point where the machine switches from reverse to forward motion.
The ground slopes there.
On this slope, the low-level controller delivers a slightly higher speed than demanded.
With its inertia, the machine overshoots the cusp.
The deviation thus quickly reaches \qty{1.54}{m}.
The third path, on the contrary, is driven right after the first one, and its initial curvature strongly differs from the final curvature of the first path.
Since the steering angle cannot change instantaneously, the machine cannot absorb such a curvature jump and temporarily strays from the path, so that the maximum path deviation reaches \qty{1.91}{m}.
The agent nevertheless adapts well to this situation and its decisions remain coherent throughout the path: after each deviation, it steers the machine back toward the reference path, and it finally brings the machine to the hopper, where it stops \qty{0.75}{m} away from the last reference point.
The completion times are close to the durations of the references, i.e., \qty{45.7}{s}, \qty{43.1}{s} and \qty{41.4}{s}.
On paths 1 and 2, they are comparable to the RL Pareto front in simulation of Fig.~\ref{fig:pareto_comparison}.
The small gap comes from the very uneven ground of the site.

\begin{table}[tb]
    \centering
    \caption{Nominal policy alone on the machine: metrics of each path}
    \label{tab:drl_machine_metrics}
    \footnotesize
    \begin{tabular}{@{}l c c c@{}}
        \toprule
        \textbf{Metric} & \textbf{Path 1} & \textbf{Path 2} & \textbf{Path 3} \\
        \midrule
        Maximum path deviation (\unit{\meter}) & \num{0.87} & \num{1.54} & \num{1.91} \\
        Mean tracking error (\unit{\meter})    & \num{0.78} & \num{1.07} & \num{1.44} \\
        Completion time (\unit{\second})        & \num{43.6} & \num{46.3} & \num{44.2} \\
        Reference duration (\unit{\second})     & \num{45.7} & \num{43.1} & \num{41.4} \\
        \bottomrule
    \end{tabular}
\end{table}

\section{Enforcing constraints with a predictive safety filter}
    \label{sec:highlevel_controller}

Section~\ref{sec:highlevelnominal} has shown that the high-level nominal policy enables the real machine to track the reference trajectories accurately.
The performance is however limited, since the environment and the machine keep changing, and an RL policy cannot by itself guarantee that the constraints are satisfied \cite{osinenko2022rlformalguarentee}.
These constraints must therefore be enforced explicitly to ensure the safe operation of the machine.
To this end, this section presents three formulations of a predictive safety filter, derived from \cite{wabersisch2018psf} and \cite{bejarano2023multistepPSF}, which filter the commands of the nominal policy and guarantee that the constraints are satisfied over the prediction horizon.

    \subsection{Filter formulation}
        \label{sec:psf_formulation}

At time step $k$, we denote by $x_{i|k}$ and $r_{i|k}$ the state and the high-level command predicted $i$ steps ahead, and by $\mathbf{x}_k=\{x_{i|k}\}_{i=0}^{H}$ and $\mathbf{r}_k=\{r_{i|k}\}_{i=0}^{H-1}$ the predicted sequences, $H$ being the prediction horizon.
All the filters introduced in this section share the same constraints.
Let $\mathcal{F}(x_k)$ be the set of pairs $(\mathbf{x}_k,\mathbf{r}_k)$ that satisfy
\begin{subequations}
    \label{eq:psf_constraints}
    \begin{align}
        \label{eq:psf_init}
        & x_{0|k} = x_k,\\
        \label{eq:psf_dynamics}
        & x_{i+1|k} = g(x_{i|k}, r_{i|k}),\quad i = 0,\dots,H-1,\\
        \label{eq:psf_state_bounds}
        & x_{i|k} \in \mathcal{Z},\quad i = 1,\dots,H,\\
        \label{eq:psf_command_bounds}
        & r_{i|k} \in \mathcal{R}_{\mathrm{c}},\quad i = 0,\dots,H-1,\\
        \label{eq:psf_clearance_constraint}
        & c_{o,b}(x_{i|k}) \ge 0,\quad i = 0,\dots,H,\ \forall o \in \mathcal{O},\ b \in \{\mathrm{f},\mathrm{r}\},\\
        \label{eq:psf_terminal}
        & x_{H|k} \in \mathcal{S}_{\mathrm{t}},
    \end{align}
\end{subequations}
The prediction starts from the measured state $x_k$ in \eqref{eq:psf_init} and follows in \eqref{eq:psf_dynamics} an analytic model of the low-level closed loop described in Section~\ref{sec:PredictionModel}.
The predicted states and high-level commands must, as stated in \eqref{eq:psf_state_bounds} and \eqref{eq:psf_command_bounds}, remain in the sets $\mathcal{Z} = \{x\colon x_{\min} \leq x \leq x_{\max} \}$ and $\mathcal{R}_{\mathrm{c}} = \{r\colon r_{\min} \leq r \leq r_{\max} \}$, respectively.
Finally, let $\mathcal{O}$ be the set of all obstacles.
As stated by \eqref{eq:psf_clearance_constraint}, the clearance $c_{o,b}$ between each body $b \in \{\mathrm{f},\mathrm{r}\}$ and each obstacle $o \in \mathcal{O}$ must remain positive over the whole horizon.
The states satisfying both \eqref{eq:psf_state_bounds} and \eqref{eq:psf_clearance_constraint} form the set $\mathcal{X}$ of Section~\ref{sec:problem_statement}.
The last predicted state must lie in the terminal safe set $\mathcal{S}_{\mathrm{t}}$, as stated in \eqref{eq:psf_terminal}.
It ensures that the machine can stop before reaching any obstacle.
The last two safety constraints are detailed in Section~\ref{sec:psf_inequality}.

Each filter then solves
\begin{equation}
    \min_{(\mathbf{x}_k,\mathbf{r}_k) \in \mathcal{F}(x_k)} J(\mathbf{x}_k,\mathbf{r}_k) + J_{\Delta}(\mathbf{r}_k),
    \label{eq:psf_generic}
\end{equation}
and applies the first command of the solution, $r_k = r^\star_{0|k}$ to the low-level controller.
The current solution is stored as the current plan and warm-starts the solve of \eqref{eq:psf_generic} at the next time step.
The filters only differ by their tracking cost $J$ penalizing the deviation to the plan proposed by the nominal policy as mentioned in Section~\ref{sec:highlevelnominal}.
The regularization
\begin{equation}
    J_{\Delta}(\mathbf{r}_k) = \lVert r_{0|k} - r_{k-1} \rVert_{W_0}^{2} + \sum_{i=1}^{H-1} \lVert r_{i|k} - r_{i-1|k} \rVert_{W_\Delta}^{2}
    \label{eq:psf_regularization}
\end{equation}
is common to all filters, including the single-step one \cite{wabersisch2018psf}.
Therefore, they are compared under the same regularization.
Here, $r_{k-1}$ is the command applied at the previous time step.
The weights $W_0$ and $W_\Delta$ penalize, respectively, the jump between two consecutive applied commands and the variations between consecutive ones of the predicted sequence, directly addressing the abrupt command changes of the nominal policy.

    \subsection{Prediction model}
        \label{sec:PredictionModel}

The PSF predicts the evolution of the machine over its horizon at every time step.
The simulator of Section~\ref{sec:simulator} is not suited to this purpose, as it involves the inference of a recurrent network.
We therefore use an analytic model, which is inexpensive to evaluate and differentiable.
Rather than reproducing every detail of the machine dynamics, it captures the dominant behavior of the closed loop, namely the second-order responses of the speed and steering low-level loops.
The kinematics $f^{\mathrm{p}}$ \eqref{eq:kinematic_model} are completed by the dynamic block $\dot{d} = f^{\mathrm{d}}(d, r)$,
\begin{subequations}
    \label{eq:analytic_model}
    \begin{align}
        \label{eq:accel}
        \dot{v}      & = a,    & \dot{a}    & = -2\xi_v\omega_v a - \omega_v^{2}(v - v^{\mathrm{c}}),\\
        \label{eq:steer_rate}
        \dot{\gamma} & = \chi, & \dot{\chi} & = -2\xi_\gamma\omega_\gamma\chi - \omega_\gamma^{2}\gamma + \omega_\gamma^{2}\gamma^{\mathrm{c}}.
    \end{align}
\end{subequations}
The speed loop requires the longitudinal acceleration $a$, thus augmenting the state $x$ of Section~\ref{sec:problem_statement}.
Throughout this section, $x$ denotes this augmented state.
In practice, $a_k$ is obtained by finite difference on the measured speed, $a_k = (v_k - v_{k-1})/\Delta t$ with $\Delta t$ being the sampling period.
The map $g$ used by the filter in constraint \eqref{eq:psf_dynamics},
\begin{equation}
    x_{k + 1} = g\left(x_k, r_k\right),
    \label{eq:map_analytic}
\end{equation}
is obtained by integrating $F = \begin{bmatrix} {f^{\mathrm{p}}}^{\top} & {f^{\mathrm{d}}}^{\top} \end{bmatrix}^{\top}$ with a fourth-order Runge--Kutta scheme, $r_k$ being held constant over each control period.
The parameters of $F$, listed in Table~\ref{tab:analytic_params}, are tuned with Bayesian optimization to match the closed-loop behavior of the real machine.
The model assumes a flat ground while, in practice, ground irregularities create discrepancies between the predicted and measured poses.
We cover these discrepancies by a safety margin $\mu$ described in Section~\ref{sec:psf_inequality}.

\begin{figure}[tb]
    \centering
    \includegraphics[width = .9\columnwidth]{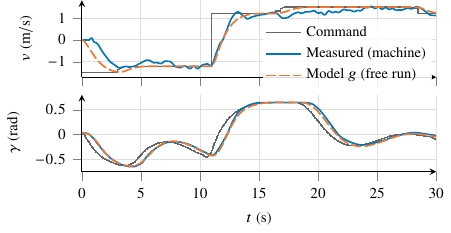}
    \caption{Free run of the analytic model $g$ compared with the real machine.}
    \label{fig:second_order}
\end{figure}

Fig.~\ref{fig:second_order} compares $g$ with the real machine where the low-level is driven by the controller of Section~\ref{sec:lowlevel_rl}.
Both receive the same commands over \qty{30}{\second}.
The model is initialized on the first measured state.
It then runs freely, without any correction from the measurements.
For this comparison, however, $g$ runs at \qty{100}{\hertz}, and not at the filter's sampling rate.
The speed and the steering angle predicted by $g$ remain close to the measurements.
In particular, $g$ reproduces the delay of the speed and steering angle response.

\begin{table}[t]
    \centering
    \caption{Parameters of the analytic full-state model $F$}
    \label{tab:analytic_params}
    \footnotesize
    \begin{tabular}{@{}l l @{\hspace{2em}} l l@{}}
        \toprule
        \textbf{Symbol} & \textbf{Value} & \textbf{Symbol} & \textbf{Value} \\
        \midrule
        $\omega_v$          & \qty{1.7435}{\radian\per\second} & $\omega_\gamma$ & \qty{1.91}{\radian\per\second} \\
        $\xi_v$             & \num{1.2261}                     & $\xi_\gamma$    & \num{0.58} \\
        \bottomrule
    \end{tabular}
\end{table}

    \subsection{Safety constraints}
        \label{sec:psf_inequality}

            \paragraph*{Obstacle constraints}

Each body $b \in \{\mathrm{f},\mathrm{r}\}$ is enclosed in a disc of radius $\rho_b$, chosen so that the disc covers the whole body.
The center $q_b$ of this disc lies on the longitudinal axis of the body, at a distance $\ell_b$ from the steering joint, so that it follows directly from the pose of the body predicted by $g$.
Each obstacle $o \in \mathcal{O}$ is enclosed in a disc of center $q_o$ and radius $\rho_{\mathrm{s}}$, which is enlarged by a model-mismatch margin $\mu$, covering the discrepancy between $g$ and the real machine.
We estimate it from data recorded on the machine.
From each measured state, $g$ is rolled out in open loop with the logged commands over $M \ge H$ steps, followed by the complete stop of the machine under $r^{\mathrm{b}}$ (backup command defined below).
The margin $\mu$ is then the largest distance between the predicted and measured disc centers reconstructed from the GPS position, the heading and the steering angle over all runs and bodies.
The clearance between body $b$ and obstacle $o$ is then
\begin{equation}
    c_{o,b}(x) = \sqrt{\lVert q_b - q_o \rVert^2 + \epsilon} - \left( \rho_b + \rho_{\mathrm{s}} + \mu \right)
    \label{eq:psf_clearance}
\end{equation}
where the small constant $\epsilon > 0$ keeps the distance differentiable when the two centers coincide.

            \paragraph*{Terminal safe set}

From the terminal safe set of \eqref{eq:psf_terminal}, the machine must be able to stop before reaching an obstacle.
The stop uses the backup command $r^{\mathrm{b}} = \begin{bmatrix} 0 & \gamma^{\mathrm{c}}_{H-1|k} \end{bmatrix}^{\top}$ setting the desired longitudinal speed to 0 and holding the last steering command of the plan.
We therefore need an upper bound on the distance travelled by the machine under $r^{\mathrm{b}}$.
Let $v^{\mathrm{b}}(t)$ and $a^{\mathrm{b}}(t) = \dot{v}^{\mathrm{b}}(t)$ be the speed and acceleration of the machine under $r^{\mathrm{b}}$, with $t = 0$ when $r^{\mathrm{b}}$ is applied.
By \eqref{eq:accel}, they follow
\begin{equation}
    \dot{a}^{\mathrm{b}} + 2\xi_v\omega_v a^{\mathrm{b}} + \omega_v^{2} v^{\mathrm{b}} = 0.
    \label{eq:speed_loop_stop}
\end{equation}
The initial conditions $v^{\mathrm{b}}(0) = v$ and $a^{\mathrm{b}}(0) = a$ are the speed and acceleration of the state $x$ from which the machine stops.
Since $\xi_v > 0$ and $\omega_v > 0$, $v^{\mathrm{b}}$ and $a^{\mathrm{b}}$ vanish as $t \to \infty$.
Integrating \eqref{eq:speed_loop_stop} over $[0, \infty)$ gives the net displacement along the path
\begin{equation}
    \int_{0}^{\infty} v^{\mathrm{b}}(t)\mathrm{d}t = \frac{2\xi_v\omega_v v + a}{\omega_v^{2}}.
    \label{eq:net_displacement}
\end{equation}
This is the stopping distance only if the speed keeps its sign.
Since the speed loop is overdamped ($\xi_v > 1$), the speed changes sign at most once, and we bound the excursion by
\begin{equation}
    D(v, a) = \frac{\lvert 2\xi_v\omega_v v + a \rvert}{\omega_v^{2}} + t_{\mathrm{d}} \lvert v \rvert .
    \label{eq:stopping_bound}
\end{equation}
The first term is the magnitude of the net displacement \eqref{eq:net_displacement}.
It is exact when the speed keeps its sign and reduces to $\lvert a \rvert / \omega_v^2$ at standstill.
The second term covers the extra forward excursion when the speed changes sign.
We checked numerically, on the discrete map $g$ (RK4, $\Delta t = \qty{0.25}{s}$), that $D$ bounds the forward and backward excursions for every direction of $(v, a)$ as soon as $t_{\mathrm{d}} \geq \qty{0.187}{s}$.
We take $t_{\mathrm{d}} = \Delta t$, also accounting for one sampling period of delay before the stop is applied.

The bound $D$ only accounts for the motion along the path.
When $r^{\mathrm{b}}$ is applied, the steering angle may still be moving towards $\gamma^{\mathrm{c}}_{H-1|k}$, which rotates the bodies around the joint and moves the disc centers even at standstill.
The steering angle and the steering command are both bounded by $\gamma_{\max}$ in absolute value, the steering angle thus varies by at most $2\gamma_{\max}$, the resulting displacement of the disc centers is bounded by the chord of a rotation of angle $2\gamma_{\max}$ and radius $L_r$
\begin{equation}
    D_\gamma = L_r \sqrt{2\left(1-\cos(2\gamma_{\max})\right)} = 2L_r \sin\gamma_{\max}.
    \label{eq:steer_bound}
\end{equation}

The terminal safe set is then
\begin{equation}
    \mathcal{S}_{\mathrm{t}} = \left\{x\colon c_{o,b}(x) \geq D(v, a) + D_\gamma,\ \forall o \in \mathcal{O},\ b \in \{\mathrm{f},\mathrm{r}\} \right\} \cap \mathcal{Z}.
    \label{eq:terminal_set}
\end{equation}
In the filter, the absolute values of \eqref{eq:stopping_bound} are calculated as $\sqrt{(\cdot)^2 + \epsilon}$, with $\epsilon > 0$, keeping the constraint smooth and only enlarging the margin.

\begin{theorem}[Recursive Safety]
    The set of constraints $\mathcal{F}(x_k)$ of \eqref{eq:psf_constraints} keeps the machine safe at every time step.
\end{theorem}

\begin{proof}
    We call a safe plan a sequence of at most $H$ high-level commands followed by the backup command $r^{\mathrm{b}}$, whose prediction by $g$ keeps $c_{o,b} \ge 0$ at every step, and assume that the machine starts in the terminal safe set, $x_0 \in \mathcal{S}_{\mathrm{t}}$.
    At $k = 0$, the backup command is a safe plan as under $r^{\mathrm{b}}$, the disc centers move by at most $D(v, a) + D_\gamma$, which is smaller than the clearance by construction of $\mathcal{S}_{\mathrm{t}}$ in \eqref{eq:terminal_set}.
    Assume now that a safe plan exists at step $k$.
    If the filter has a solution at time $k+1$, this solution satisfies $c_{o,b} \ge 0$ at every predicted step and ends in $x_{H|k+1} \in \mathcal{S}_{\mathrm{t}}$, from which the same argument as for $k = 0$ holds to stop under $r^{\mathrm{b}}$.
    Extended by $r^{\mathrm{b}}$, the solution is therefore a safe plan, replacing the stored one.
    Otherwise, the filter applies the next command of the stored plan and the remainder of this plan, still followed by $r^{\mathrm{b}}$, is a safe plan at step $k + 1$.
    By induction, the machine is safe at every time step (as long as the mismatch between $g$ and the machine remains within the margin $\mu$).
\end{proof}

    \subsection{Tracking costs}
        \label{sec:psf_articulated}

The three filters we compare in this article differ by their tracking cost measuring the deviation from the nominal plan: the cost $J$ of \eqref{eq:psf_generic} is replaced by $J_{\mathrm{s}}$, $J_{\mathrm{m}}$ or $J_{\mathrm{e}}$, defined below for the single-step, multi-step and enhanced multi-step filters, respectively.
This plan is obtained by rolling out the nominal policy $\pi_{\mathrm{hl}}$ of Section~\ref{sec:highlevelnominal} in the filter's model, i.e.
\begin{equation}
    \begin{aligned}
        x^{\mathrm{ext}}_{0|k} & = x_k, & r^{\mathrm{ext}}_{i|k}   & = \pi_{\mathrm{hl}}\left(x^{\mathrm{ext}}_{i|k}, \mathcal{P}_{i|k}, r^{\mathrm{ext}}_{i-1|k}\right), \\
                               &        & x^{\mathrm{ext}}_{i+1|k} & = g\left(x^{\mathrm{ext}}_{i|k}, r^{\mathrm{ext}}_{i|k}\right)
    \end{aligned}
    \label{eq:psf_rollout}
\end{equation}
where $\mathcal{P}_{i|k}$ is the window of the path seen by the controller at predicted step $i$, and $r^{\mathrm{ext}}_{-1|k}=r_{k-1}$ as $\pi_{\mathrm{hl}}$ is recurrent, its internal state is propagated along the rollout from a copy of its current value.
Therefore, the prediction does not alter the policy actually running on the machine.
This rollout provides the proposed commands $\mathbf{r}^{\mathrm{ext}}_k = \{r^{\mathrm{ext}}_{i|k}\}_{i=0}^{H-1}$ and, at no extra cost, the state $\mathbf{x}^{\mathrm{ext}}_k = \{x^{\mathrm{ext}}_{i|k}\}_{i=0}^{H-1}$.

The single-step PSF (SPSF) \cite{wabersisch2018psf,tearle2021predictive} only filters the first nominal command $r^{\mathrm{ext}}_{0|k}$
\begin{equation}
    J_{\mathrm{s}}(\mathbf{x}_k,\mathbf{r}_k) = \lVert r_{0|k} - r^{\mathrm{ext}}_{0|k} \rVert_{W}^{2}.
    \label{eq:psf_single}
\end{equation}
The rest of the horizon only certifying that a safe continuation exists.
The correction is thus chosen without regard to the commands that follow, which can cause chattering and large-magnitude corrections \cite{bejarano2023multistepPSF}.
Following \cite{bejarano2023multistepPSF}, we also consider a multi-step PSF (MPSF), which penalizes the deviation from the nominal commands over the whole horizon
\begin{equation}
    J_{\mathrm{m}}(\mathbf{x}_k,\mathbf{r}_k) = \sum_{i=0}^{H-1} \lVert r_{i|k} - r^{\mathrm{ext}}_{i|k} \rVert_{W}^{2}
    \label{eq:psf_multi}
\end{equation}
Both formulations only compare high-level commands.
After a correction, the filtered trajectory may drift away from the nominal one while its commands remain close to the nominal ones, delaying the return to the intended path.
We therefore also penalize the deviation of the predicted state $x_{i|k}$ from the nominal state $x^{\mathrm{ext}}_{i|k}$ on both bodies, since on an articulated machine the rear body does not follow the path of the front one.
Dropping the index k for readability, the errors $e^{\mathrm{lon}}_{b,i}$, $e^{\mathrm{lat}}_{b,i}$, $e^{\psi}_{b,i}$ of body $b \in \{\mathrm{f},\mathrm{r}\}$ at predicted step $i$ are computed as in \eqref{eq:hl_elon}, \eqref{eq:hl_elat} and \eqref{eq:hl_epsi} with $\Delta X_{b,i} = X_{b,i} - X^{\mathrm{ext}}_{b,i}$ and $\Delta Y_{b,i} = Y_{b,i} - Y^{\mathrm{ext}}_{b,i}$.
Expressed in the body's frame, they separate a lag along the path from a lateral deviation, which can be weighted independently.
These errors are gathered in
\begin{equation}
    e_i = \Big[
        e^{\mathrm{lon}}_{\mathrm{f},i}\ \ e^{\mathrm{lat}}_{\mathrm{f},i}\ \ e^{\psi}_{\mathrm{f},i}\ \ e^{\mathrm{lon}}_{\mathrm{r},i}\ \ e^{\mathrm{lat}}_{\mathrm{r},i}\ \ e^{\psi}_{\mathrm{r},i}\ \ v_i - v^{\mathrm{ext}}_i
    \Big]^{\top}.
    \label{eq:psf_error_vector}
\end{equation}
From this, we propose an enhanced multi-step PSF (EPSF) in which the tracking cost blends, at each predicted step, a command term and a state term
\begin{equation}
    J_{\mathrm{e}}(\mathbf{x},\mathbf{r}) = \sum_{i=0}^{H-1} \left[ \tau_i \lVert r_i - r^{\mathrm{ext}}_i \rVert_{W}^{2} + (1-\tau_i) \lVert e_i \rVert_{Q}^{2} \right],
    \label{eq:psf_articulated}
\end{equation}
with $Q$ a positive semi-definite diagonal matrix and the weighting factor
\begin{equation}
    \tau_i = \exp\left(-\alpha \sqrt{\lVert e_i \rVert^2 + \epsilon_\tau}\right) \in (0, 1]
    \label{eq:psf_tau}
\end{equation}
where $\alpha \ge 0$ sets the sensitivity of the blend to the state deviation and the small constant $\epsilon_\tau > 0$ keeps $\tau_i$ differentiable at zero deviation.
As long as the filtered state stays close to the nominal one, $\tau_i \approx 1$ and the filter essentially tracks the nominal commands, as in \eqref{eq:psf_multi}.
When a constraint has pushed it away, $\tau_i \to 0$ and the cost switches to bringing the state back onto the nominal trajectory rather than chasing commands that no longer fit the diverged state.
Setting $\alpha = 0$ gives $\tau_i = 1$ and recovers \eqref{eq:psf_multi}.
As $\tau_i$ depends on the decision variable $x_i$, $J_{\mathrm{e}}$ remains smooth but is not a quadratic form with fixed weights.
As for the headings, the steering entries of $W$, $W_0$ and $W_\Delta$ are expressed per squared degree, i.e., each command weight $\bar{W}$ enters the cost as $S\bar{W}S$ with $S = \mathrm{diag}(1, \kappa_{\deg})$, so that all the angular terms of the cost share the same scale.

Finally, since $\mathbf{x}^{\mathrm{ext}}_k$ is the rollout \eqref{eq:psf_rollout} of $\mathbf{r}^{\mathrm{ext}}_k$ through the same map $g$, the nominal plan $(\mathbf{x}^{\mathrm{ext}}_k,\mathbf{r}^{\mathrm{ext}}_k)$ sets all the tracking costs to zero.
Hence, when this plan is feasible and $W_0 = W_\Delta = 0$, it is a minimizer of \eqref{eq:psf_generic} and passes through the filter unchanged: the state term of $J_{\mathrm{e}}$ does not turn the filter into a tracking MPC, it only shapes the correction when one is needed.
With $W_0, W_\Delta \neq 0$, the filter also smooths a feasible plan, which is a deliberate departure from a purely minimal correction.
Note that feasibility includes the terminal constraint so that near an obstacle, the stopping distance at the end of the nominal plan may exceed the clearance.
The tail of the plan is then bent even when its first commands are safe.

\section{Validation of the full architecture}
    \label{sec:results}

In this section, the three PSFs are first compared in simulation under identical conditions, and the best one is then deployed on the machine.

    \subsection{Simulation results}

\begin{figure}[tb]
    \centering
    \subfloat[Trajectories]{\includegraphics[width = 0.6\linewidth]{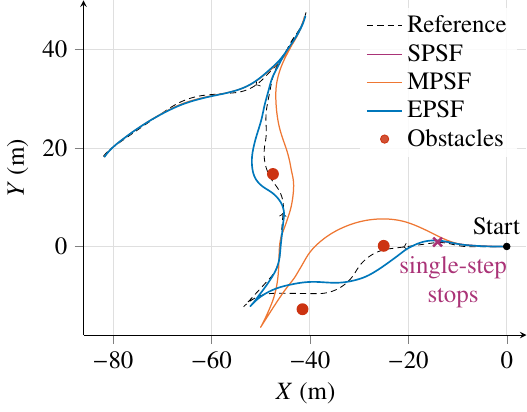}\label{fig:psf_traj}}\\
    \subfloat[SPSF]{\includegraphics[width = 0.9\linewidth]{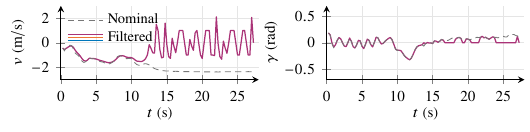}\label{fig:psf_cmd_single}}\\
    \subfloat[MPSF]{\includegraphics[width = 0.9\linewidth]{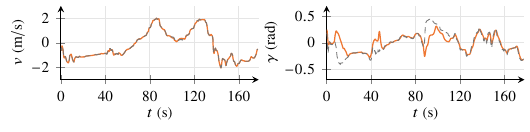}\label{fig:psf_cmd_multistep}}\\
    \subfloat[EPSF]{\includegraphics[width = 0.9\linewidth]{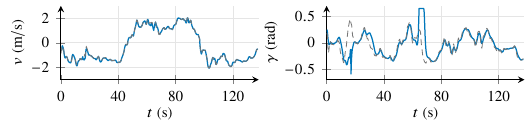}\label{fig:psf_cmd_state_tracking}}
    \caption{Simulation results of the PSF obstacle avoidance scenario.}
    \label{fig:psf_obstacle_avoidance}
\end{figure}

In simulation, we consider a complex and long trajectory with several obstacles on it.
The obstacles are squares of \qty{1}{m} by \qty{1}{m}, each enclosed in its circumscribed disc, as described in Section~\ref{sec:psf_inequality}.
The parameters of the filters were tuned on a different scenario, which is representative of the target application but more demanding than the one used for the evaluation.
The results of the simulations are presented on Fig~\ref{fig:psf_obstacle_avoidance}.
As expected, the SPSF \eqref{eq:psf_single} performs worse than the other filters.
By correcting only $r_{0|k}$, it causes strong chattering near the first obstacle and the machine never manages to get around it.
Such chattering can damage the actuators and the transmission of the machine.
The SPSF is therefore not suited to a real deployment.
As reported in the literature \cite{bejarano2023multistepPSF} and seen in Fig.~\ref{fig:psf_cmd_multistep}, the MPSF reduces this chattering significantly, allowing the machine to navigate around obstacles more effectively.
Nonetheless, it strongly deviates from the trajectory and this behavior is undesirable for industrial applications.
The EPSF \eqref{eq:psf_articulated} addresses this drawback.
It goes around every obstacle, like the MPSF.
Unlike the MPSF, it then brings the machine back onto the reference path.
The state term of its cost penalizes the deviation of both bodies from the nominal plan.
So that once an obstacle is passed, the filtered trajectory is driven back to the nominal one.
The machine therefore follows the reference trajectory closely, except in the vicinity of the obstacles.
Moreover, its commands show no chattering as presented in Fig.~\ref{fig:psf_cmd_state_tracking}.
The speed and steering commands remain smooth, including during the safety corrections.
As for the MPSF, this smoothness comes from the rate penalties $W_0$ and $W_\Delta$.
The EPSF thus combines safety, tracking precision, and smooth commands.

The metrics of the three filters are given in Table~\ref{tab:psf_simu_results}.
The speed and steering commands are evaluated separately.
The intervention of the filter is the mean absolute difference between the applied command $r_k$ and the command $r^{\mathrm{ext}}_{0|k}$ proposed by the nominal policy.
The chattering is measured by the mean rate of change of the commands, $|v^{\mathrm{c}}_k - v^{\mathrm{c}}_{k-1}|/\Delta t$ and $|\gamma^{\mathrm{c}}_k - \gamma^{\mathrm{c}}_{k-1}|/\Delta t$, with $\Delta t = \qty{0.25}{s}$.
The minimum clearance is the smallest distance between a body and an obstacle over the run.
Finally, as on the machine, the completion time is measured from the first reference point until the machine stops at the end of the path.
It is not reported for the SPSF, since the machine starts chattering in front of the first obstacle at $t=\qty{27}{s}$ and never reaches the end of the path.

\begin{table}[tb]
    \centering
    \caption{Simulation results of the PSF obstacle avoidance scenario}
    \label{tab:psf_simu_results}
    \footnotesize
    \setlength{\tabcolsep}{4pt}
    \begin{tabular}{@{}l c c c@{}}
        \toprule
        & \textbf{SPSF} & \textbf{MPSF} & \textbf{EPSF} \\
        \midrule
        Mean tracking error (\unit{\meter})                & \num{1.65}$^{*}$ & \num{3.60}     & \textbf{1.49} \\
        Mean speed intervention (\unit{\meter\per\second}) & \num{1.304}      & \textbf{0.010} & \num{0.022} \\
        Mean steering intervention (\unit{\radian})        & \textbf{0.027}   & \num{0.072}    & \num{0.082} \\
        Minimum obstacle clearance (\unit{\meter})         & \num{0.167}      & \textbf{1.587} & \num{1.079} \\
        Mean speed rate (\unit{\meter\per\second\squared}) & \num{2.040}      & \textbf{0.136} & \num{0.229} \\
        Mean steering rate (\unit{\radian\per\second})     & \num{0.147}      & \textbf{0.037} & \num{0.070} \\
        Completion time (\unit{\second})                   & --               & \num{177.3}    & \textbf{137.3} \\
        \bottomrule
    \end{tabular}

    \vspace{2pt}

    {\scriptsize $^{*}$Computed over the \qty{27}{s} before the machine starts chattering in front of the first obstacle.}
\end{table}

    \subsection{Machine results}

\begin{figure}[tb]
    \centering
    \includegraphics[width = 0.9\columnwidth]{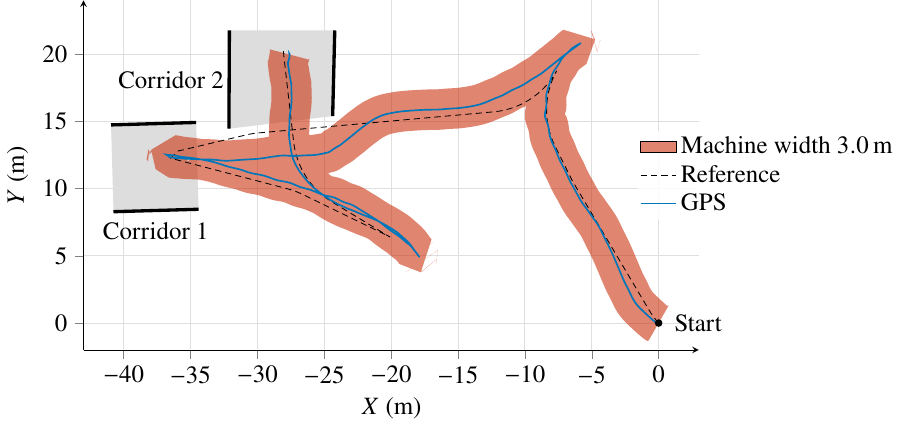}
    \caption{EPSF deployed on the machine.}
    \label{fig:psf_machine_corridors}
\end{figure}

The machine performs a production cycle passing through corridors in which the loader is forced to move within narrow bounds.
These corridors are located at usual material loading and unloading areas.
We deploy the best of the three filters evaluated in simulation, i.e., the EPSF.
This scenario challenges the machine less than the simulation one as, under industrial conditions, the paths are normally planned so that the machine moves safely in the corridors.
We therefore add a difficulty: the reference path is placed close to the edge of corridor 2.
The scenario is shown in Fig.~\ref{fig:psf_machine_corridors}.
The reference path passes \qty{0.75}{m} from the walls of corridors where the machine performs an avoidance maneuver.
The distance between the GPS antenna and the walls is never below \qty{2.04}{m}.
Since the antenna is mounted at the center of the machine in the lateral direction and the total width of the machine is \qty{3.012}{m}, the side of the machine therefore always stays at least about \qty{0.5}{m} from the walls of corridors.
Once these walls are passed, the machine returns onto the reference.

The inference time is measured by the period between two successive filtered commands sent to the machine.
This period includes the wait for the next high-level step, since a new PSF solve is triggered at \qty{4}{Hz}, as stated in Section~\ref{sec:architecture}.
The computation of the next path at the cusp is excluded.
The mean inference time is \qty{0.28}{s}, slightly above the nominal period of \qty{0.25}{s} because each cycle of the scheduler that triggers the high-level controller at \qty{4}{Hz} adds a small overhead.
The PSF keeps up with this rate throughout the run, including during the avoidance maneuver.
The only exceedance, up to \qty{0.59}{s}, occurs at the end of the run, while the machine slowly approaches the final point.
This is consistent with the expected rate, showing that the PSF computation takes one period at nearly every step.

\section{Conclusion}
    \label{sec:conclusion}

This article addresses the constrained trajectory tracking of a weakly instrumented articulated wheeled loader with a hierarchical architecture.
At the low-level, a reinforcement learning (RL) agent tracks the speed commands, ensuring that the machine follows the desired velocity profile.
At the high-level, an EPSF corrects the commands of an RL nominal policy using an inexpensive analytic model of the low-level closed loop.
Deployed without any fine-tuning, the low-level agent tracks the speed more accurately than PID controllers on the real machine.
In simulation, the EPSF avoids every obstacle and, by penalizing the deviation of both bodies from the nominal plan, significantly reduces the mean tracking error compared with the MPSF.
On the real machine, it keeps the loader about \qty{0.5}{m} away from the walls of narrow corridors and returns it onto the reference, while running at the \qty{4}{Hz} rate of the high-level controller.
Future work will exploit the repetitive nature of the work cycles to improve the tracking performance from one iteration to the next.
The prediction model of the filter will also include time-varying parameters, to better capture the changes of the low-level closed loop with the operating conditions.
Finally, the high-level controller will also be extended with convergence and stability guarantees \cite{millios2024stabilitypsf, didier2021adaptivempsf}.

\bibliographystyle{IEEEtran}
\bibliography{2026_roullier_l_tcst}

@article{azulay2021rlwheelloader,
    title   = {Wheel Loader Scooping Controller Using Deep Reinforcement Learning},
    author  = {Azulay, Osher and Shapiro, Amir},
    journal = {IEEE Access},
    volume  = {9},
    number  = {},
    pages   = {24145-24154},
    year    = {2021},
    doi     = {10.1109/ACCESS.2021.3056625}
}

@inproceedings{bejarano2023multistepPSF,
    title     = {Multi-step model predictive safety filters: reducing chattering by increasing the prediction horizon},
    author    = {Bejarano, Federico Pizarro and Brunke, Lukas and Schoellig, Angela P.},
    booktitle = {Proc. 62nd IEEE Conf. Decis. Control},
    pages     = {4723--4730},
    year      = {2023}
}

@article{borquez2024hjr,
    title   = {On Safety and Liveness Filtering Using {H}amilton–{J}acobi Reachability Analysis},
    author  = {Borquez, Javier and Chakraborty, Kaustav and Wang, Hao and Bansal, Somil},
    journal = {IEEE Trans. Robotics},
    volume  = {40},
    number  = {},
    pages   = {4235-4251},
    year    = {2024},
    doi     = {10.1109/TRO.2024.3454470}
}

@article{dadhich2016surveywheelloader,
    title   = {Key challenges in automation of earth-moving machines},
    author  = {S. Dadhich and U. Bodin and U. Andersson},
    journal = {Autom. Constr.},
    volume  = {68},
    pages   = {212-222},
    year    = {2016},
    doi     = {10.1016/j.autcon.2016.05.009}
}

@article{egli2021rlexcavatorclassic,
    title   = {A General Approach for the Automation of Hydraulic Excavator Arms Using Reinforcement Learning},
    author  = {Egli, Pascal and Hutter, Marco},
    journal = {IEEE Robot. Autom. Lett.},
    volume  = {7},
    number  = {2},
    pages   = {5679-5686},
    year    = {2022},
    doi     = {10.1109/LRA.2022.3152865}
}

@inproceedings{frank2012introwheelloader,
    title        = {On increasing fuel efficiency by operator assistant systems in a wheel loader},
    author       = {Frank, Bobbie and Skogh, L. and Filla, R. and Alak{\"{u}}la, Mats},
    booktitle    = {Proc. Int. Conf. Adv. Veh. Technol. Integr.},
    pages        = {155--161},
    year         = {2012},
    doi          = {10.13140/RG.2.1.3129.1362}
}

@misc{guerrier2024learningcontrolbarrierfunctions,
    title         = {Learning Control Barrier Functions and their application in Reinforcement Learning: A Survey},
    author        = {Maeva Guerrier and Hassan Fouad and Giovanni Beltrame},
    year          = {2024},
    doi           = {10.48550/arXiv.2404.16879},
    url           = {https://arxiv.org/abs/2404.16879},
    archiveprefix = {arXiv},
    eprint        = {2404.16879},
    primaryclass  = {cs.LG}
}

@article{ishiara2020excavatorMPC,
    title   = {Realization of Excavator Loading Operation by Nonlinear Model Predictive Control with Bucket Load Estimation},
    author  = {Shinji Ishihara and Akira Kanazawa and Ryu Narikawa},
    journal = {IFAC Pap. OnLine},
    volume  = {54},
    number  = {20},
    pages   = {20-25},
    year    = {2021},
    doi     = {https://doi.org/10.1016/j.ifacol.2021.11.147},
}

@inproceedings{millios2024stabilitypsf,
    title     = {Stability Mechanisms for Predictive Safety Filters},
    author    = {Milios, Elias and Wabersich, Kim Peter and Berkel, Felix and Schwenkel, Lukas},
    booktitle = {Proc. 63rd IEEE Conf. Decis. Control},
    pages     = {2409-2416},
    year      = {2024},
    doi       = {10.1109/CDC56724.2024.10886293}
}

@inproceedings{nayl2012wheelloadermodel,
    title     = {Switching model predictive control for an articulated vehicle under varying slip angle},
    author    = {Nayl, Thaker and Nikolakopoulos, George and Gustfsson, Thomas},
    booktitle = {Proc. 20th Mediterr. Conf. Control Automation},
    volume    = {},
    number    = {},
    pages     = {890-895},
    year      = {2012},
    doi       = {10.1109/MED.2012.6265751}
}

@article{nguyen2023surveywheelloader,
    title   = {Robotic autonomous systems for earthmoving equipment operating in volatile conditions and teaming capacity: a survey},
    author  = {Nguyen, Huynh A.D. and Ha, Quang P.},
    journal = {Robotica},
    volume  = {41},
    number  = {2},
    pages   = {486–510},
    year    = {2023},
    doi     = {10.1017/S0263574722000339}
}

@inproceedings{Penttila2025wheelloadermpcrl,
    title     = {Autonomous Wheel Loader Navigation Using Goal-Conditioned Actor-Critic MPC},
    author    = {Mäki-Penttilä, Aleksi and Toulkani, Naeim Ebrahimi and Ghabcheloo, Reza},
    booktitle = {Proc. IEEE Int. Conf. Robot. Autom.},
    volume    = {},
    number    = {},
    pages     = {1286-1292},
    year      = {2025},
    doi       = {10.1109/ICRA55743.2025.11127375}
}

@misc{Schulman2017PPO,
    title         = {Proximal Policy Optimization Algorithms},
    author        = {John Schulman and Filip Wolski and Prafulla Dhariwal and Alec Radford and Oleg Klimov},
    year          = {2017},
    doi           = {10.48550/arXiv.1707.06347},
    url           = {https://arxiv.org/abs/1707.06347},
    archiveprefix = {arXiv},
    eprint        = {1707.06347},
    primaryclass  = {cs.LG},
}

@article{tearle2021predictive,
    title     = {A predictive safety filter for learning-based racing control},
    author    = {Tearle, Ben and Wabersich, Kim P. and Carron, Andrea and Zeilinger, Melanie N.},
    journal   = {IEEE Robot. Autom. Lett.},
    volume    = {6},
    number    = {4},
    pages     = {7635--7642},
    year      = {2021},
    publisher = {IEEE}
}

@article{wabersisch2018psf,
    title   = {A predictive safety filter for learning-based control of constrained nonlinear dynamical systems},
    author  = {Kim Peter Wabersich and Melanie N. Zeilinger},
    journal = {Automatica},
    volume  = {129},
    pages   = {109597},
    year    = {2021},
    doi     = {10.1016/j.automatica.2021.109597}
}

@article{wabersisch2023surveysaferl,
    title   = {Data-Driven Safety Filters: Hamilton-{J}acobi Reachability, Control Barrier Functions, and Predictive Methods for Uncertain Systems},
    author  = {Wabersich, Kim P. and Taylor, Andrew J. and Choi, Jason J. and Sreenath, Koushil and Tomlin, Claire J. and Ames, Aaron D. and Zeilinger, Melanie N.},
    journal = {IEEE Control Syst. Mag.},
    volume  = {43},
    number  = {5},
    pages   = {137-177},
    year    = {2023},
    doi     = {10.1109/MCS.2023.3291885}
}

@article{feng2025newsurveywheelloader,
    title = {Advancements and challenges in earthmoving equipment automation using core technologies},
    author = {Jiyu Feng and Wei Chen and Changyi Liu and Xinhui Liu and Bingwei Cao and Feixiang Xu},
    journal = {Autom. Constr.},
    volume = {179},
    pages = {106472},
    year = {2025},
    doi = {10.1016/j.autcon.2025.106472},
}

@article{zhang2020analyticmodelwheelloader,
    title = {Using gyro stabilizer for active anti-rollover control of articulated wheeled loader vehicles},
    author = {Zhang, Xiaoyu and Liu, Qianjie and Liu, Jun and Zhu, Qingyuan and Hu, Huosheng},
    journal = {Proc. Inst. Mech. Eng. Part I J. Syst. Control Eng.},
    volume = {235},
    pages = {095965182093711},
    year = {2020},
    doi = {10.1177/0959651820937114}
}

@article{kumar2024operatormodelwheeledloader,
    title = {Operator model for wheel loader short-cycle loading handling},
    author = {Manoranjan Kumar and Torbjörn Ekevid and Welf Löwe},
    journal = {Autom. Constr.},
    volume = {167},
    pages = {105691},
    year = {2024},
    doi = {10.1016/j.autcon.2024.105691},
}

@article{aoshima2024sim2realwheelloader,
    title = {Examining the simulation-to-reality gap of a wheel loader digging in deformable terrain},
    author = {Aoshima, Koji and Servin, Martin},
    journal = {Multibody Syst. Dyn.},
    volume = {64},
    pages = {121-148},
    year = {2024},
    doi = {10.1007/s11044-024-10005-5}
}

@inproceedings{spinelli2024rlhydraulicmaterialhandling,
    title={Reinforcement Learning Control for Autonomous Hydraulic Material Handling Machines with Underactuated Tools}, 
    author={Spinelli, Filippo A. and Egli, Pascal and Nubert, Julian and Nan, Fang and Bleumer, Thilo and Goegler, Patrick and Brockes, Stephan and Hofmann, Ferdinand and Hutter, Marco},
    booktitle={Proc. IEEE Int. Conf. Intell. Robots Syst}, 
    year={2024},
    volume={},
    pages={12694-12701},
    doi={10.1109/IROS58592.2024.10802199}
}

@inproceedings{hoffmann2025lstmexcavator,
    title={Modeling Weakly-Instrumented Excavator Arm Dynamics with Stacked-Input LSTM},
    author={Hoffmann, Nicolas
    and Cohen, Max
    and Preda, Marius
    and Zaharia, Titus},
    booktitle={Robot. Intell. Technol. Appl.},
    year={2025},
    pages={206--218},
    doi={10.1007/978-3-031-92011-0_17}
}

@article{victoria2021bayesianoptimizationhp,
    title = {Automatic tuning of hyperparameters using {B}ayesian optimization},
    author = {Victoria, Helen and Maragatham, Ganesh},
    journal = {Evol. Syst.},
    year = {2021},
    volume = {12},
    doi = {10.1007/s12530-020-09345-2}
}

@inproceedings{asmar2023mppiadaptiveimportancesampling,
    title={Model Predictive Optimized Path Integral Strategies}, 
    author={Asmar, Dylan M. and Senanayake, Ransalu and Manuel, Shawn and Kochenderfer, Mykel J.},
    booktitle={Proc. IEEE Int. Conf. Robot. Autom.},
    year={2023},
    pages={3182-3188},
    doi={10.1109/ICRA48891.2023.10160929}
}

@inproceedings{keshavarz2025mppice,
    title={Control of Legged Robots Using Model Predictive Optimized Path Integral}, 
    author={Keshavarz, Hossein and Ramirez-Serrano, Alejandro and Khadiv, Majid},
    booktitle={Proc. 2025 IEEE-RAS 24th Int. Conf. Humanoid Robots}, 
    year={2025},
    pages={1-8},
    doi={10.1109/Humanoids65713.2025.11203053}
}

@inproceedings{majd20254mppitractor,
    title={{GPU}-Accelerated Barrier-Rate Guided {MPPI} Control for Tractor-Trailer Systems}, 
    author={Majd, Keyvan and Parwana, Hardik and Hoxha, Bardh and Hong, Steven and Okamoto, Hideki and Fainekos, Georgios},
    booktitle={Proc. 28th IEEE Trans. Intell. Transp. Syst.}, 
    year={2025},
    pages={3451-3458},
    doi={10.1109/ITSC60802.2025.11423534}
}

@article{lahr2026l4acados,
    title={L4acados: Learning-Based Models for acados, {A}pplied to {G}aussian Process-Based Predictive Control}, 
    author={Lahr, Amon and Näf, Joshua and Wabersich, Kim P. and Frey, Jonathan and Siehl, Pascal and Carron, Andrea and Diehl, Moritz and Zeilinger, Melanie N.},
    journal={IEEE Trans. Control Syst. Technol.}, 
    year={2026},
    volume={34},
    pages={1091-1105},
    doi={10.1109/TCST.2025.3638421}
}

@inproceedings{andersson2021rlforestrycrane,
    title={Reinforcement Learning Control of a Forestry Crane Manipulator}, 
    author={Andersson, Jennifer and Bodin, Kenneth and Lindmark, Daniel and Servin, Martin and Wallin, Erik},
    booktitle={Proc. IEEE Int. Conf. Intell. Robots Syst.}, 
    year={2021},
    pages={2121-2126},
    doi={10.1109/IROS51168.2021.9636219}
}

@inproceedings{roullier2024excavatorcontrolcomparison,
    title={Data-Driven Control of a Weakly-Instrumented Excavator with Deep Learning}, 
    author={Hoffmann, N. and Cohen, M. and Roullier, L. and Preda, M. and Zaharia, T.},
    booktitle={Proc. 3rd IEEE Int. Conf. Intell. Real.}, 
    year={2024},
    pages={1-8},
    doi={10.1109/ICIR64558.2024.10976848}
}

@article{wang2022surveycurriculumlearning,
    title={A Survey on Curriculum Learning}, 
    author={Wang, Xin and Chen, Yudong and Zhu, Wenwu},
    journal={IEEE Trans. Pattern Anal. Mach. Intell.}, 
    year={2022},
    volume={44},
    number={9},
    pages={4555-4576},
    doi={10.1109/TPAMI.2021.3069908}
}

@article{sun2024mpcdrlcommandcorrection,
    title={A Novel Framework Combining {MPC} and Deep Reinforcement Learning With Application to Freeway Traffic Control}, 
    author={Sun, Dingshan and Jamshidnejad, Anahita and De Schutter, Bart},
    journal={IEEE Trans. Intell. Transp. Syst.}, 
    year={2024},
    volume={25},
    number={7},
    pages={6756-6769},
    doi={10.1109/TITS.2023.3342651}
}

@article{zhang2025mpcdrlmodelcorrection,
    title = {Model predictive control with residual learning and real-time disturbance rejection: Design and experimentation},
    author = {Haodi Zhang and Junwei Ge and Jinya Su and Kun Gu and Fuyou Wang and Wen-Hua Chen and Shihua Li},
    journal = {Control Eng. Pract.},
    year = {2025},
    volume = {165},
    pages = {106587},
    doi = {https://doi.org/10.1016/j.conengprac.2025.106587}
}

@inproceedings{hoffmann2007stanley,
    title={Autonomous Automobile Trajectory Tracking for Off-Road Driving: Controller Design, Experimental Validation and Racing},
    author={Hoffmann, Gabriel M. and Tomlin, Claire J. and Montemerlo, Michael and Thrun, Sebastian},
    booktitle={Proc. Am. Control Conf.},
    pages={2296--2301},
    year={2007},
    doi={10.1109/ACC.2007.4282788}
}

@article{osinenko2022rlformalguarentee,
    title = {Reinforcement learning with guarantees: a review},
    author = {Pavel Osinenko and Dmitrii Dobriborsci and Wolfgang Aumer},
    journal = {IFAC-PapersOnLine},
    year = {2022},
    volume = {55},
    number = {15},
    doi = {https://doi.org/10.1016/j.ifacol.2022.07.619},
}

@book{GrunePannek2017MPC,
    title     = {Nonlinear Model Predictive Control: Theory and Algorithms},
    author    = {Gr{\"u}ne, Lars and Pannek, J{\"u}rgen},
    edition   = {2nd},
    year      = {2017},
    publisher = {Springer London},
    doi       = {10.1007/978-3-319-46024-6}
}

@inproceedings{didier2021adaptivempsf,
    title={Adaptive Model Predictive Safety Certification for Learning-based Control}, 
    author={Didier, Alexandre and Wabersich, Kim P. and Zeilinger, Melanie N.},
    booktitle={Proc.IEEE Conf. Decis. Control}, 
    year={2021},
    pages={809-815},
    doi={10.1109/CDC45484.2021.9682832}
}

\end{document}